\documentclass[10pt,journal,compsoc]{IEEEtran}
\usepackage{amsmath,amssymb,amsfonts}
\usepackage{algorithm}
\usepackage{algorithmic}
\usepackage{graphicx}
\usepackage{textcomp}

\mathchardef\mhyphen="2D
\graphicspath{ {./figures/} }
\usepackage{cleveref}
\usepackage{xcolor}
\usepackage{multirow}
\usepackage{subfigure}
\usepackage{url}
\usepackage{pifont}

\newcommand\dblcheck[1][darkgray]{\textcolor{#1}{\rlap{\;\ding{51}}\ding{51}}\;}
\newcommand\snglcheck[1][darkgray]{\textcolor{#1}{\ding{51}}}
\newcommand\xmark[1][darkgray]{\textcolor{#1}{\ding{55}}}
\newcommand*\encircle[1]{\raisebox{1.1pt}{\textcircled{\raisebox{-1pt} {#1}}}}
\newcommand\tab[1][1cm]{\hspace*{#1}}

\newcommand{\setalglineno}[1]{%
  \setcounter{ALC@line}{\numexpr#1-1}}

\newcommand\algorithmicprocedure{\textbf{procedure}}
\newcommand{\algorithmicendprocedure}{\algorithmicend\ \algorithmicprocedure}
\makeatletter
\newcommand\PROCEDURE[3][default]{%
  \ALC@it
  \algorithmicprocedure\ \textsc{#2}(#3)%
  \ALC@com{#1}%
  \begin{ALC@prc}%
}
\newcommand\ENDPROCEDURE{%
  \end{ALC@prc}%
  \ifthenelse{\boolean{ALC@noend}}{}{%
    \ALC@it\algorithmicendprocedure
  }%
}
\newenvironment{ALC@prc}{\begin{ALC@g}}{\end{ALC@g}}
\makeatother

\ifCLASSOPTIONcompsoc
  \usepackage[nocompress]{cite}
\else
  \usepackage{cite}
\fi

\usepackage{ragged2e}
\newcommand*\rot{\rotatebox{-60}}

\newcommand*\rottt{\rotatebox{-30}}
\usepackage{array}
\newcolumntype{P}[1]{>{\centering\arraybackslash}m{#1}}
\newcolumntype{R}[1]{>{\raggedleft\arraybackslash}p{#1}}
\newcolumntype{L}[1]{>{\raggedright\arraybackslash}m{#1}}
\newcolumntype{M}[1]{>{\centering\arraybackslash}m{#1}}

\begin{document}
\title{An Ensemble Mobile-Cloud Computing Method for Affordable and Accurate Glucometer Readout}

\author{Navidreza Asadi, and Maziar Goudarzi,~\IEEEmembership{Senior Member, IEEE}
\thanks{N. Asadi is now with Computer Engineering Department, Technical University of Munich, Germany (navidreza.asadi@tum.de). He was with Computer Engineering Department, Sharif University of Technology, Iran while working on this project.}
\thanks{M. Goudarzi is with the Computer Engineering Department, Sharif University of Technology, Iran (goudarzi@sharif.edu).}
\thanks{Manuscript submitted to IEEE Transactions on Mobile Computing.}
}

\markboth{Pre-print. Submitted to IEEE Transactions on Mobile Computing.}%
{N. Asadi \MakeLowercase{\textit{et al.}}: Ensemble-based Mobile-Cloud Approach for Image-Based Logging From Glucometers}

\IEEEtitleabstractindextext{%

\begin{justify}\begin{abstract}
Despite essential efforts towards advanced wireless medical devices for regular monitoring of blood properties, many such devices are not available or not affordable for everyone in many countries. 
Alternatively using ordinary devices, patients ought to log data into a mobile health-monitoring manually. 
According to medical specialists, it causes several issues: 
(1)~due to the direct human intervention, it is prone to errors, and clients reportedly tend to enter unrealistic data; 
(2)~typing values several times a day is bothersome and causes clients to leave the mobile app. Thus, there is a strong need to use now-ubiquitous smartphones, reducing error by capturing images from the screen of medical devices and extracting useful information automatically. Nevertheless, there are a few challenges in its development: 
(1)~data scarcity has led to impractical methods with very low accuracy: to our knowledge, only small datasets are available in this case; 
(2)~accuracy-availability tradeoff: one can execute a less accurate algorithm on a mobile phone to maintain higher availability, or alternatively deploy a more accurate and more compute-intensive algorithm on the cloud, however, at the cost of lower availability in poor/no connectivity situations. We present an ensemble learning algorithm, a mobile-cloud computing service architecture, and a simple compression technique to achieve higher availability and faster response time while providing higher accuracy by integrating cloud- and mobile-side predictions. Additionally, we propose an algorithm to generate synthetic training data which facilitates utilizing deep learning models to improve accuracy. Our proposed method achieves three main objectives: 
(1)~$92.1\%$ and $97.7\%$ accuracy on two different datasets, improving previous methods by ${\sim}40\%$, 
(2)~reducing required bandwidth by $45{\times}$ with  ${\sim}1\%$ drop in accuracy, 
(3)~and providing better availability compared to
 mobile-only, cloud-only, split computing, and early exit service models.
\end{abstract}\end{justify}

}

\maketitle

\IEEEdisplaynontitleabstractindextext
\IEEEpeerreviewmaketitle

\IEEEraisesectionheading{\section{Introduction}\label{sec:introduction}}
\IEEEPARstart{M}{any} mHealth/uHealth medical devices, especially those affordable in middle/low-income countries, show the measured quantity on a digital or seven-segment screen alongside other additional information such as date, time, diagrams and measurement units. 
In most commonly used mHealth services, particularly for diabetics, patients are required to manually type sensed values into their mobile app.
As illustrated in Fig.\,\ref{fig:logging_manually}, a client first reads a value from the medical device, opens the app, navigates to logging interface, and eventually logs the information through typing. These steps should be repeated every time they log a measurement.

\subsection{Motivation}

According to the reports \cite{Given2013, Salvi2020} as well as our own experience from our diabetes management application iDia \cite{idia}, this procedure has a few drawbacks:
(1)~Manual logging multiple times a day, is deterring; it is bothersome and time-consuming, and based on the feedbacks we have received, leads users to eventually lose their interest in using the app.
(2)~It is prone to human errors.
More importantly, the observations show that patients are tempted to enter fake information that are more acceptable and closer to the normal values. This can have considerable negative effects on the whole process of prevention, control, and treatment.

\begin{figure*}[t!]
    \centering
    \subfigure[Manual]
    {
        \includegraphics[width=0.3\textwidth]{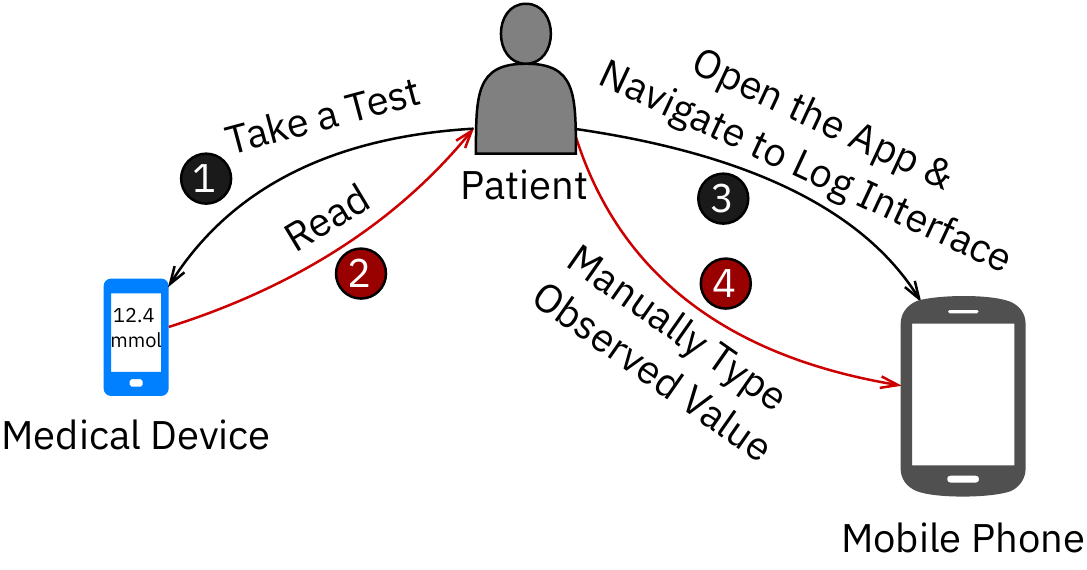}
        \label{fig:logging_manually}
    }
    ~
    \subfigure[Wireless Communication]
    {
        \includegraphics[width=0.3\textwidth]{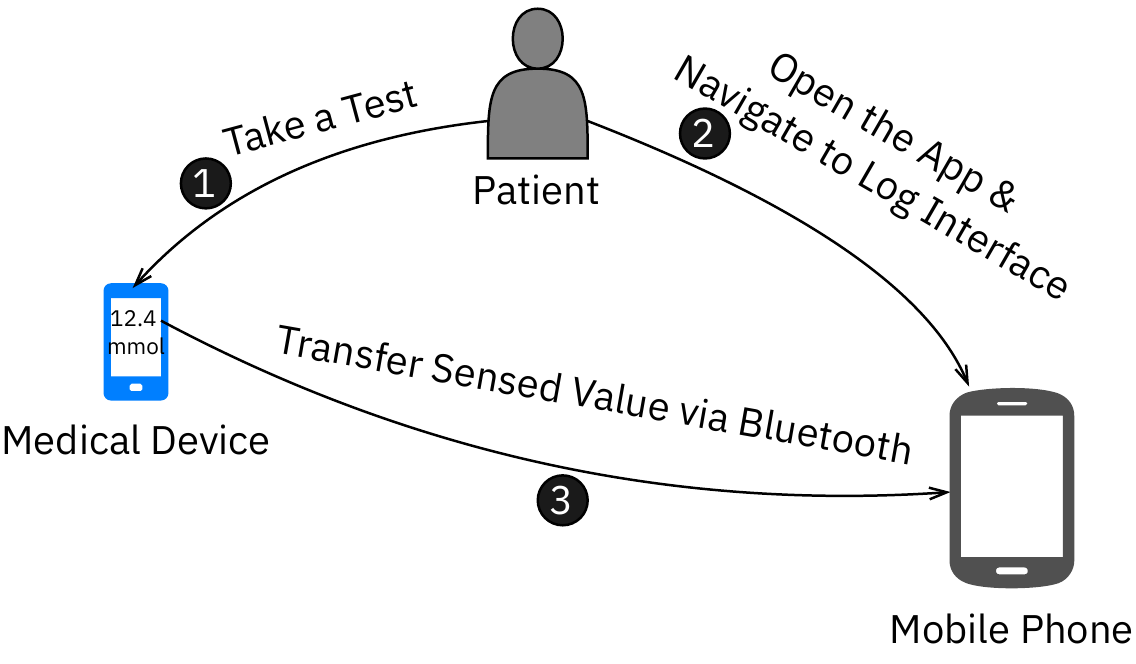}
        \label{fig:logging_bluetooth}
    }
    ~
    \subfigure[Image-Based]
    {
        \includegraphics[width=0.3\textwidth]{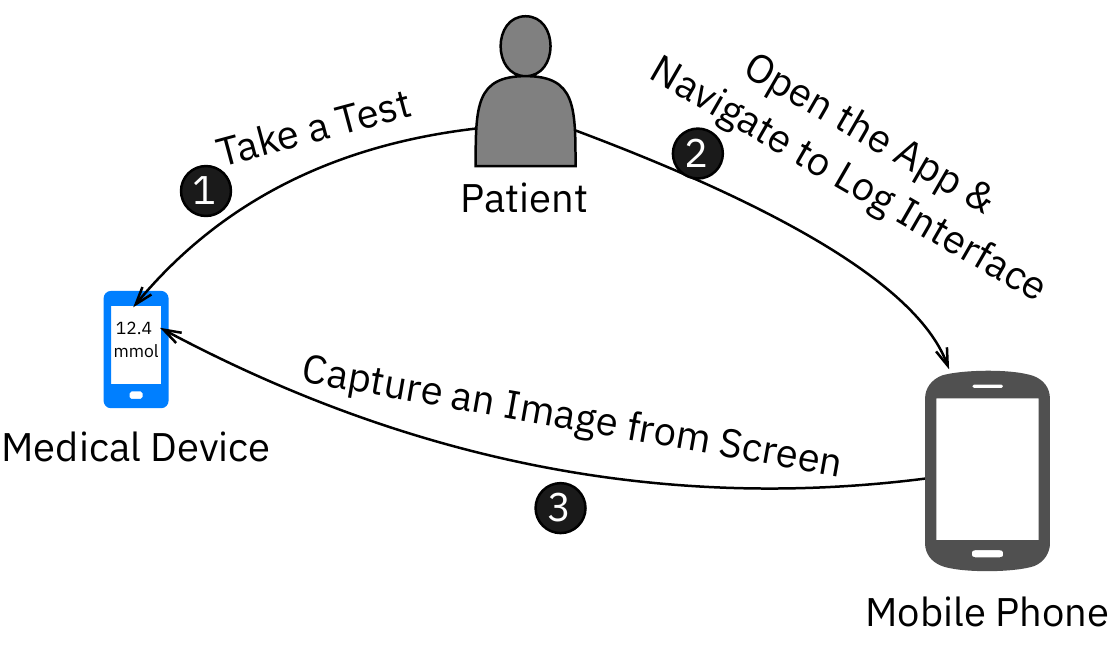}
        \label{fig:logging_imagecapturing}
    }
	\vskip -0.15in
    \caption{Logging methods. Removing human interventions (red arrows)  is preferred to avoid false logs.}
    \label{fig:logging_methods}
	\vskip -0.15in
\end{figure*}

\begin{figure}[t!]
  \centerline{\includegraphics[width=0.75\columnwidth]{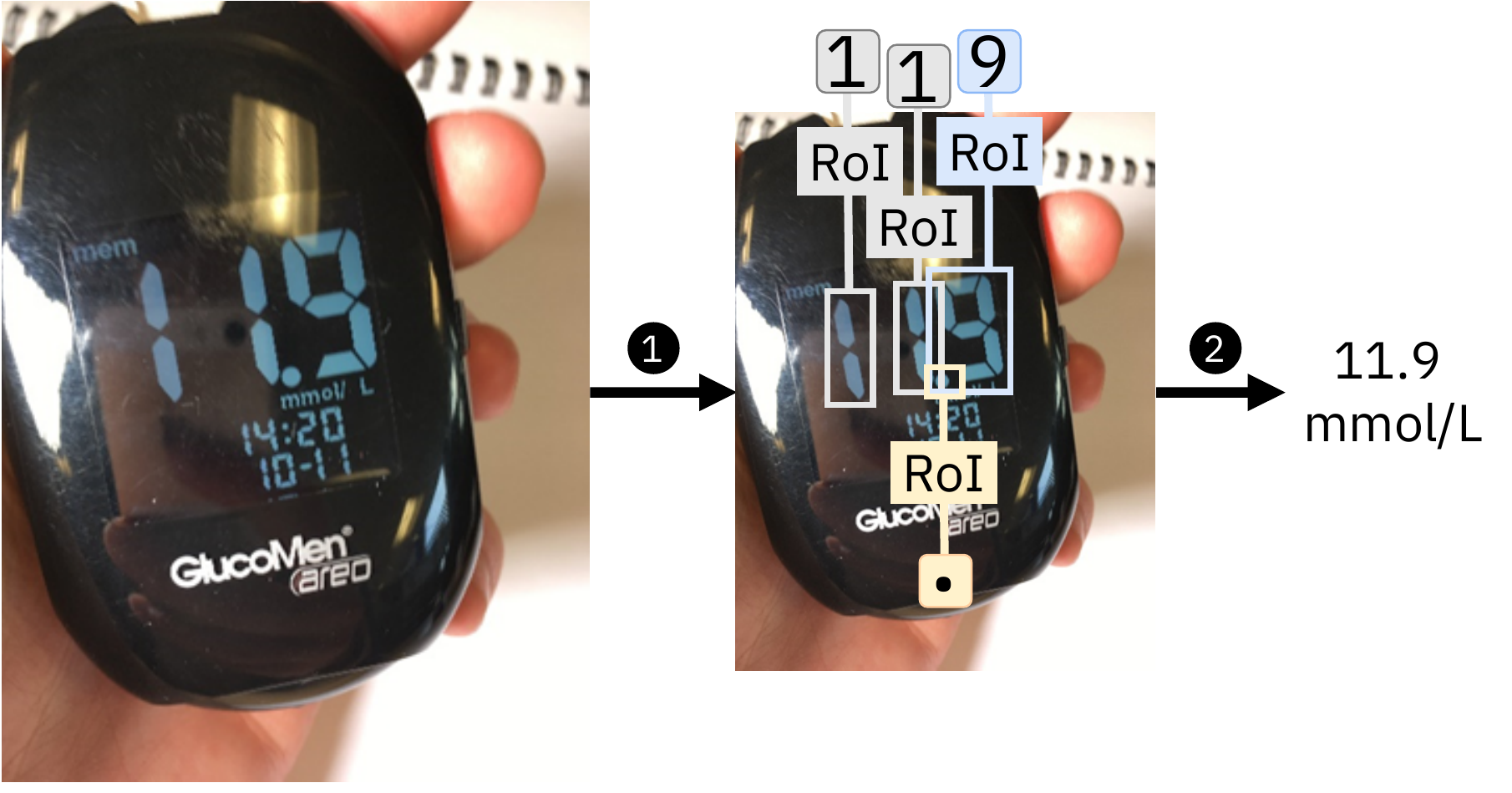}}
  \vskip -0.15in
  \caption{Image-based method, showing Regions-of-Interest (RoIs) and the correct labels. The expected correct readout is ``11.9 mmol/L''.}
  \label{fig:image_based_method}
  \vskip -0.15in

\end{figure}

Currently, there are two better approaches:
(1) Some medical devices are able to transmit their data to mobile devices via distant communication technologies (e.g., Bluetooth), facilitating the logging procedure (Fig.\,\ref{fig:logging_bluetooth}). Nevertheless, they are far more expensive and in some cases less accurate
\cite{continuous_glucose_monitoring}. More importantly, most of them can only interact with their own software applications and do not permit third-party apps to receive data. In addition, different versions of transmission technologies cause incompatibility between different mobile and medical devices.
Despite a potentially bright future, these devices currently hold less than one percent of the market \cite{bmeterguide}. This number is even lower in developing countries.   
(2) An alternative which has grown interest in academia, uses image capturing and computing capabilities of mobile devices (i.e., smartphones); they are  available to almost everybody and are able to perform light-weight computing tasks. Fig.\,\ref{fig:logging_imagecapturing} illustrates this alternative: the mobile phone is used to capture an image of the medical device, and then image processing is applied to recognize the sensed values automatically. In this approach (Fig.\,\ref{fig:image_based_method}), each digit is considered as an object having a region of interest (RoI) and an ordered sequence of digits (i.e., RoIs) forms a \textit{read} string. A correct read means not only all digits are predicted correctly, but also in the same order as displayed on the device.
In this paper we follow this methodology for its broader applicability especially in middle/low-income countries.

  \subsection{Challenges}
  
    \textbf{Accuracy.}	Although automated image-based reading may seem basic and easy, there are  a few points that prove it a challenging problem in this specific case: (1) the current state-of-the-art \cite{finnegan2019automated}, which has largely improved the previous algorithms, uses only two medical devices, yet achieves just \textbf{51.5\%} accuracy, meaning it misreads almost half of the captured images. 
    (2) as we discuss in Sections \ref{sec_related_work} and \ref{sec_proposed_method}, the diversity of medical devices and a variety of structural and visual differences as well as noisy information on their screens (e.g., date and time) make it extremely difficult to read sensed value with business-as-usual image processing techniques.
    \newline
    \textbf{Data Scarcity.}	The accuracy of deep learning algorithms relies on either big volume of annotated data to be trained in a supervised manner or a model pre-trained on a related task. To the best of our knowledge, there exists neither a big dataset nor a related pre-trained model in our particular task of imaged-based reading. Generative models (e.g., GANs) also would not help there because they require similar data to train, which is not applicable in our case. Additionally, it might be really difficult (if not impossible) to generate automatic annotation for our usecase.   
    \newline
    \textbf{Resource Constraints.} Current state-of-the-art deep learning models are compute-intensive and need to be deployed on specialized cloud infrastructures. Thus, they introduce new challenges, including availability during poor network conditions, that is genuinely an expected issue in our target under-developed countries.
    Edge Computing \cite{Shi2016}, referring to every device near data source with some compute capacity (e.g., smartphones), is considered as a promising approach to improve availability and quality of service (QoS) by reducing delay. Mobile devices are usually resource-constrained, and therefore, can only run simpler deep neural network (DNN) models with much lower accuracy. Cloud Computing, however, is at the opposite side.
  
  \subsection{Main Contributions}
\begin{figure}
    \centerline{\includegraphics[width=\columnwidth]{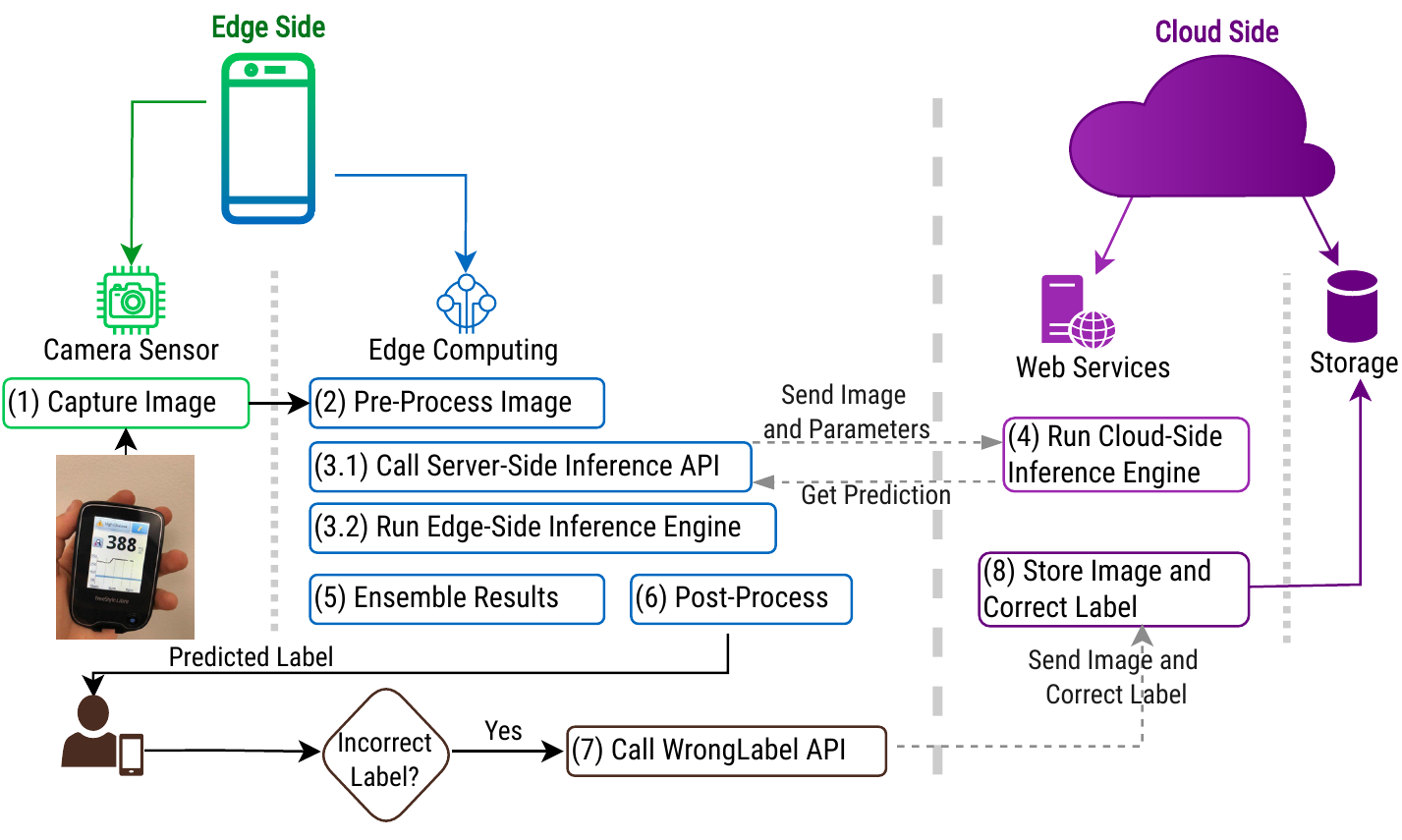}}
    \vskip -0.15in
    \caption{Our mobile-cloud architecture.}
    \vskip -0.15in
    \label{fig:mobile_cloud_architecture}
  \end{figure}

    We propose practical solutions to address the challenges above. We present an ensemble mobile-cloud computing architecture to get the best of both worlds: higher availability on the mobile, as well as higher accuracy and enhanced performability (i.e., a measure of level of performance/service-quality of the system) by integrating cloud and mobile modules (Fig.~\ref{fig:mobile_cloud_architecture}, described in \S\ref{sec_proposed_method}). The followings items are our main contributions:
\newline 
(1) A hybrid mobile-cloud service architecture, together with a compatible ensemble deep learning algorithm. 
This enables an accuracy-availability tradeoff based on network connectivity. We addressed the challenges of combining predictions of two separate models; e.g., differences and overlaps in the identified bounding boxes for each data element.
\newline 
(2) A simple yet effective compression method.
Combined with our ensemble model, this provides higher accuracy despite little communicated data.
\newline 
(3) A high-fidelity data synthesizer algorithm, making utilizing deep learning models possible.
This has basically turned the challenge into an opportunity; the variety of glucometer models, data formats, units, fonts, etc. is a challenge to conventional methods, but we used it in our data synthesizer mechanism to produce enough reasonable data to train high accuracy models that cover many varieties including those not seen before.

Our proposed method achieves $92.1\%$ and $97.7\%$ accuracy on two real-world datasets, and improves previously published results by more than $40\%$.
It reduces the required bandwidth by $45{\times}$, and maintains higher availability compared to mobile-only, cloud-only, split computing, and early exit service models.
Our proposal can be easily extended to other usecases with minor modifications.

The rest of this paper is organized as follows: In Section\,\ref{sec_related_work}, we review related work. In Section\,\ref{sec_proposed_method}, we present our proposed method. Section\,\ref{sec_data_generation} explains our dataset generation algorithm. We evaluate our methods and algorithms in Section\,\ref{sec_evaluation} and conclude in Section\,\ref{sec_conclusion}.
  
  \section{Related Work}
  \label{sec_related_work}

  \begin{figure*}[!t]
      \centerline{\includegraphics[width=.8\textwidth]{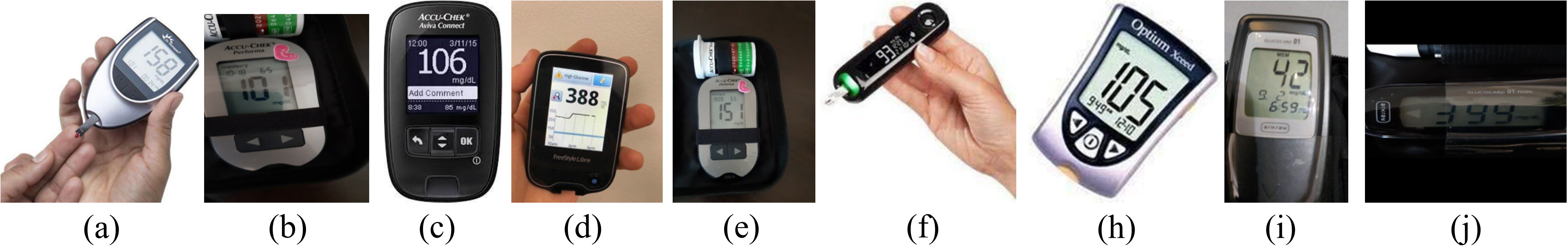}}
      \vskip -0.15in
      \caption{Examples of images captured from medical devices.}
      \label{fig:example_images}
  \end{figure*}
  
    We separate related work into two different parts. The first one presents algorithms for reading sensed values or detection and recognition of digits on digital or seven-segment screens. The second one summarizes the studies that attempt to deploy part or whole of a deep learning model on the mobile.
  
  \subsection{Image-Based Automated Reading}

  \begin{table}[b]
    \centering
    \vskip -0.2in
    \caption{Summary of Literature Review on Image-Based Automated Reading.}
    \label{tab:related_work_1}
    \setlength{\tabcolsep}{3pt}
    \vskip -0.1in
    
    \begin{tabular}{P{1cm}|R{1.15cm}R{1.15cm}R{1.15cm}R{1.15cm}R{1.15cm}}
      \hline
      \rot{Work} & \rot{Localization} & \rot{Accuracy} & \rot{Robustness*} & \rot{Generalization} & \rot{Response Time} \\
      \hline
      \cite{morris2006clearspeech} & \xmark & \xmark & \xmark & \xmark & \snglcheck \\
      \cite{Ghugardare2009} & \xmark & \xmark & \xmark & \xmark & ? \\
      \cite{Mariappan2011} & \xmark & \xmark & \xmark & \xmark & ? \\
      \cite{Ghosh2014} & \xmark & \xmark & \xmark & \xmark & ? \\
      \cite{Tekin2011} & \snglcheck & \xmark & \xmark & \xmark & \snglcheck \\
      \cite{Rasines2012} & \snglcheck & \xmark & \xmark & \xmark & \snglcheck \\
      \cite{Tesseract} & \snglcheck & \xmark & \xmark & \xmark & \snglcheck \\
      \cite{Prakruthi2017} & \xmark & \xmark & \xmark & \xmark & \xmark \\
      \hline
      \cite{Liu2016b} & \snglcheck & \xmark & \snglcheck & \snglcheck & \snglcheck \\
      \cite{finnegan2019automated} & \snglcheck & \xmark & \snglcheck & \xmark & \snglcheck \\
      \hline
      Ours & \snglcheck & \snglcheck & \snglcheck & \snglcheck & \snglcheck \\
      \hline
      \multicolumn{6}{p{200pt}}{*\- Robustness to different environmental conditions.} \\
      \multicolumn{6}{p{200pt}}{?\- Information not available.}
      \end{tabular}
    \vskip -0.3in
  \end{table}

    Most of the proposed methods break the problem into multiple steps including image enhancement, localization of RoIs, detection and classification, and eventually ordering. 
    We review related work within five criteria, as summarized in Table~\ref{tab:related_work_1}.
    \newline
    \textbf{Automated Localization.}
    Locating the RoIs is a crucial step and impacts the final accuracy.
    Many works try to simplify the problem while assuming the localization step is somehow already done, either manually by a client, or by fixing the device or using special markers \cite{morris2006clearspeech, Ghugardare2009, Mariappan2011, Ghosh2014, Rasines2012, Prakruthi2017}. A few others \cite{Tekin2011, Liu2016b, finnegan2019automated}, and our work take a more holistic approach, applying automated localization as well.
    \newline
    \textbf{Accuracy.}
    To our knowledge, no previous work achieves a reasonable readout accuracy. \cite{finnegan2019automated} with $51.5\%$ has by far the best performance. General OCR engines also are not helpful; our previous experiments along with the reports from \cite{Liu2016b, Prakruthi2017}, and \cite{finnegan2019automated} confirm the state-of-the-art OCR engine, Tesseract's \cite{Tesseract} poor performance ($<$10\%) in this specific task. 
    On the other hand, our method can reach over $90\%$ accuracy on both datasets.
    \newline
    \textbf{Robustness.}
    Different methods, especially those using conventional algorithms are usually sensitive to different conditions such as skewness, various noises, camera perspective and angle, exposure, and illumination.
    Here, we define robustness as performing consistently in different conditions. That said, only few works \cite{finnegan2019automated, Liu2016b} try to adress it. We simulate all these variations in our data synthesizer. Addtionally, the ensemble of two different models, mitigates such errors.
    \newline
    \textbf{Generalization.}
    There are a variety of medical devices each having unique characteristics, such as different font styles (including various types of seven-segments and/or digital styles), background and foreground colors, screen size and shape, units, backlit, etc. (Fig.~\ref{fig:example_images}). 
    Nevertheless, the previous studies, except \cite{Liu2016b}, use one or very few specific devices so that their proposed algorithms highly depend on the properties of those selected devices; hence, may not be considered as a general solution. For example, \cite{Rasines2012} is designed for a particular medical device with a blue backlit screen, \cite{finnegan2019automated} considers only seven-segment screens, and\cite{Prakruthi2017} assumes the largest contour as the screen, and hardcodes exact location of RoIs. In contrast, we cover a broad market, and illustrate this using an additional public dataset from Oxford \cite{finnegan2019automated}.
    \newline
    \textbf{Respone Time.}
    Since larger portion of the previous studies use conventional and light-weight image processing methods, they can achieve reasonable response time. For instance, \cite{Rasines2012} is implemented on a Samsung Galaxy S i9000 mobile device, and can process 20 frames per second, or \cite{Tekin2011} is deployed on an N95 mobile device, and achieves five frames per second.
    The only exception is \cite{Prakruthi2017} which uses a deep convolutional neural network (CNN) for the digits classification step. It takes 10 seconds, that is unsatisfactory. Although we leverage CNN-based object detection models, we choose parameters so that the response is prepared in less than half a second. Besides, we design a simple compression technique when using the cloud-side engine in poor network conditions, and, therefore, reduce the end-to-end response time.
    
  \subsection{Deep Learning on Edge}
  
  \begin{figure*}[t!]
      \centering
      \subfigure[Cloud-Only]
      {
          \includegraphics[width=0.30\textwidth]{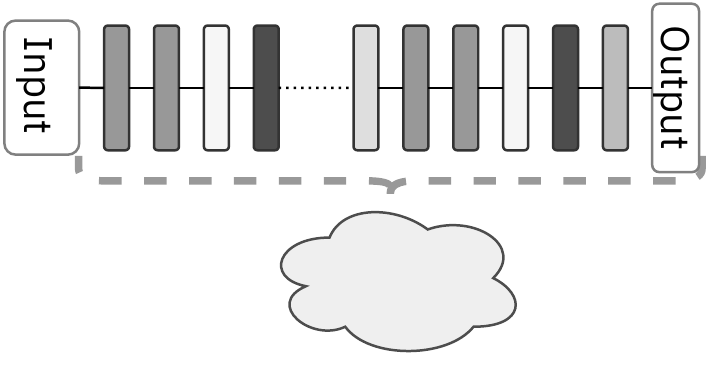}
          \label{fig:service_models_cloud_only}
      }
      ~
      \subfigure[Mobile-Only]
      {
          \includegraphics[width=0.18\textwidth]{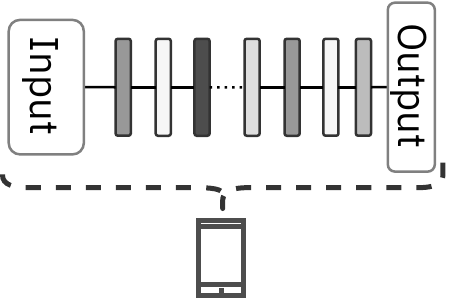}
          \label{fig:service_models_edge_only}
      }
      ~
      \subfigure[Split Computing]
      {
          \includegraphics[width=0.30\textwidth]{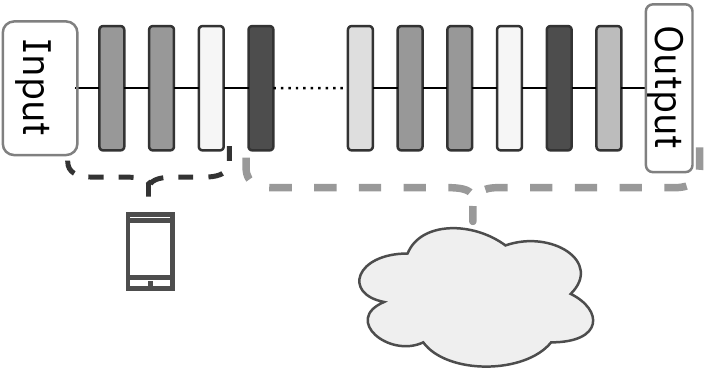}
          \label{fig:service_models_split_computing}
      }
      ~
      \subfigure[Early Exit]
      {
          \includegraphics[width=0.25\textwidth]{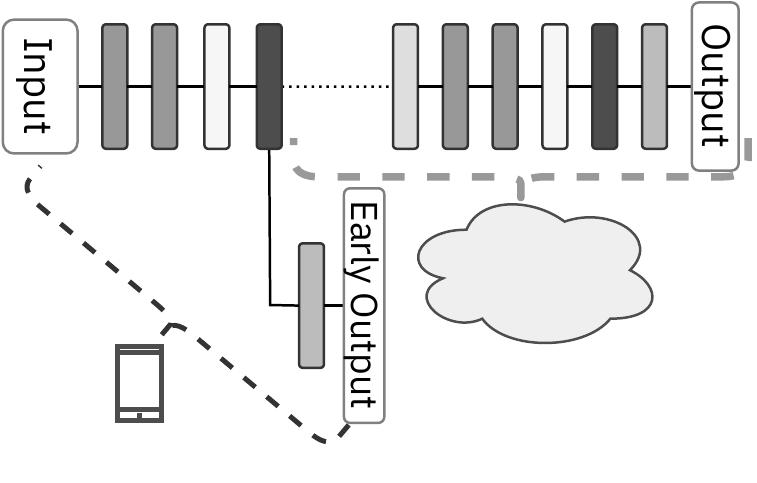}
          \label{fig:service_models_early_exit}
      }
      ~
      \subfigure[Our Architecture]
      {
          \includegraphics[width=0.5\textwidth]{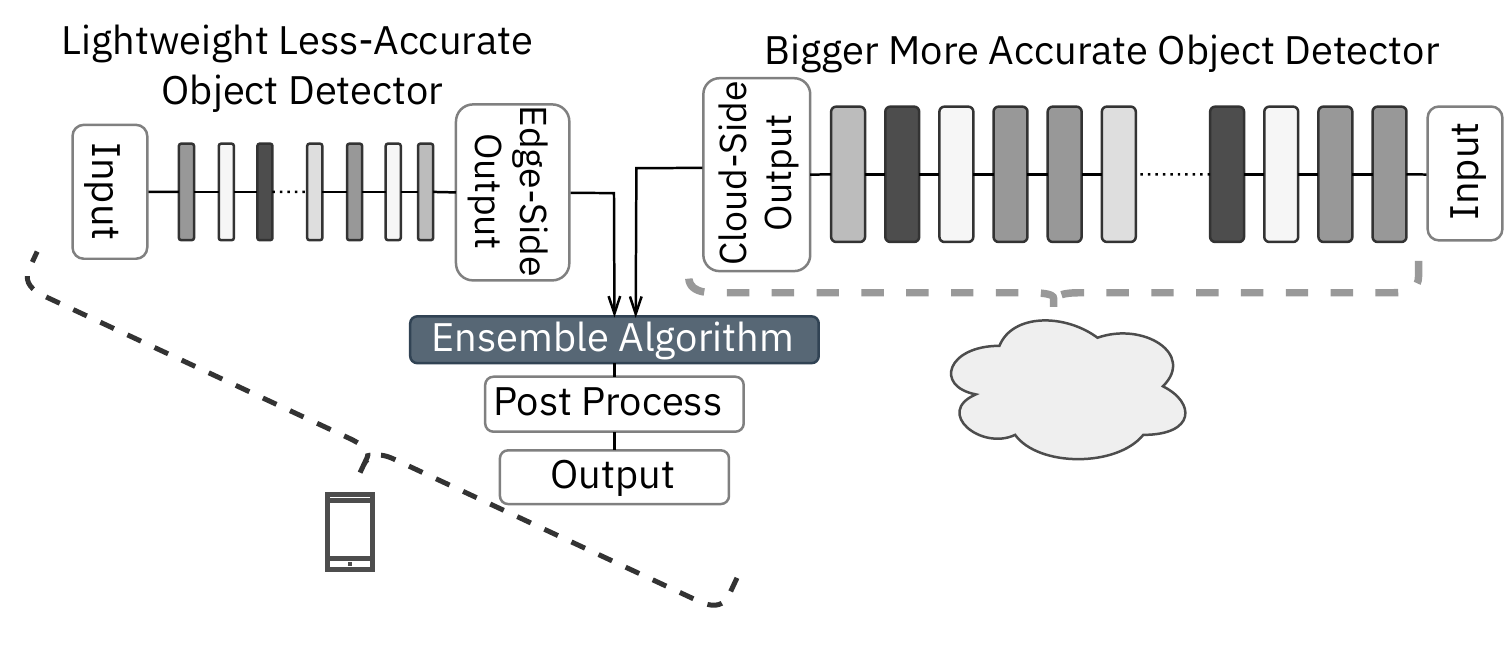}
          \label{fig:service_models_proposed}
      }
      \vskip -0.11in
      \caption{Deep Learning Serving Architectures.}
      \label{fig:serving_architectures}
    \vskip -0.2in
  \end{figure*}

	In practice, an edge device can be any computing machine (indluding smartphones) that generates data or is near a data generation source. Edge devices are usually resource-constrained. So it is challenging to deploy big deep learning models on edge. A concise comparison of different methods is provided in Table~\ref{tab:related_work_2}.
  The efforts to overcome the limitations can be divided into three major directions.

  \begin{table}[b]
    \vskip -0.2in
    \caption{Related Work on Deep Learning at Edge: A summary.}
    \label{tab:related_work_2}
    \setlength{\tabcolsep}{3pt}
    \vskip -0.1in
    \begin{tabular}{P{0.8cm}|P{1.9cm}|R{.75cm}R{.75cm}R{.7cm}R{.75cm}R{0.75cm}R{.85cm}}
      \hline
      \rot{Method} & \rot{Works} & \rot{Accuracy} & \rot{Availability} & \rot{No-Cloud*} & \rot{Bandwidth**} & \rot{Response Time} & \rot{Deployment} \\
      \hline
      {D1} & \cite{Sandler2018, Dai2019, Howard2019a, Tan2019b, Rastegari2020} & \xmark & \snglcheck & \snglcheck & - & \snglcheck & E \\
      \hline
      {D2} {A1} & \cite{Li2020c, Hsieh2018, Wang2018a} & \xmark & \snglcheck & \xmark & \snglcheck & \snglcheck & E,C \\
      \hline
      {D2} {A2} & \cite{Kang2017a, Hu2020, Laskaridis2020a} & \snglcheck & \xmark & \xmark & \snglcheck & ? & E,C \\
      \hline
      {D3} & \cite{Laskaridis2020a, Teerapittayanon2017, Zhang2019a} & \xmark & \snglcheck & \snglcheck & \xmark & \snglcheck & E or C \\
      \hline
      \multicolumn{2}{P{2.7cm}|}{Ours} & \snglcheck & \snglcheck & \snglcheck & \snglcheck & \snglcheck & E,C \\
      \hline
      \multicolumn{7}{p{200pt}}{*\- Independence from a central cloud entity.} \\
      \multicolumn{7}{p{200pt}}{**\- Bandwidth usage optimization.} \\
      \multicolumn{7}{p{200pt}}{D: Direction \quad A: Approach \quad E: Edge \quad C: Cloud}
      \end{tabular}
    \vskip -0.3in
  \end{table}

	The first direction is designing light-weight DNN models or optimizing existing ones. Several successful works have studied light-weight models including
	\cite{Sandler2018, Dai2019, Howard2019a, Tan2019b, Rastegari2020}, and \cite{Redmon2017}.
	In general, related work in this category leverage a combination of using convolution blocks with lower parameters (e.g., separable convolutions), quantization, pruning, and model distillation. These techniques mainly focus on optimizing response time and memory footprint, and thereby sacrifice accuracy (Fig.\,\ref{fig:service_models_edge_only}).

	The second direction, distributes computation across edge and central cloud, vertically (Fig.\,\ref{fig:service_models_split_computing}). Some techniques aim to reduce the required computation and bandwidth by dropping insignificant frames of input stream at the edge, before sending them to the cloud, and depending on the nature of a task may follow different filtering policies \cite{Li2020c, Hsieh2018, Wang2018a}.
	Some distribute the inference model across edge and cloud (Split Computing) \cite{Kang2017a, Hu2020, Laskaridis2020a}, and usually trade accuracy for better response time or bandwidth usage. These methods rely on central cloud; hence, in the case of network outages, they become unavailable. 

	Another recent direction determines one or more exit points within neural model (including pre-processing steps). 
  Exit points are usually designed so that at least one of them stay on the edge. Depending on the task and its requirements, the model can exit early, sacrificing accuracy to meet delay constraints	\cite{Laskaridis2020a, Teerapittayanon2017, Zhang2019a}. 
	Despite their appealing results, they still are immature and are evaluated on simple tasks such as classification (Fig.\,\ref{fig:service_models_early_exit}).

	In comparison, the strength of our proposed mobile-cloud architecture (Fig.\,\ref{fig:service_models_proposed}) is its ability to take advantage of both worlds: it improves accuracy, and in general performability as well as availability, thanks to the independent nature of our ensemble models.

\section{Proposed Method}
\label{sec_proposed_method}

  Our service captures images from medical devices using phone's camera (Fig.\,\ref{fig:mobile_cloud_architecture}(1)), then performs a pre-processing step on mobile (Fig.\,\ref{fig:mobile_cloud_architecture}(2)).
	The mobile device concurrently sends the prepared image to cloud (Fig.\,\ref{fig:mobile_cloud_architecture}(3.1, 4)), while executing light-weight inference engine locally (Fig.\,\ref{fig:mobile_cloud_architecture}(3.2)).
	After receiving predictions of both sides, the mobile device runs our ensemble algorithm (Fig.\,\ref{fig:mobile_cloud_architecture}(5)).
	It also performs a post-processing step, to correct the initial answer that is produced by the mobile, if required (Fig.\,\ref{fig:mobile_cloud_architecture}(6)). 
	If user recognizes a misprediction, they can send the image and its corresponding true reading (Fig.\,\ref{fig:mobile_cloud_architecture}(7)) to cloud storage for future analysis and training iterations (Fig.\,\ref{fig:mobile_cloud_architecture}(8)). 
	The mobile model provides availability, while the cloud model improves performability (Refer to \S\ref{subsec:availability_performability}). Both models together improve accuracy.

	We reduce the problem into object detection so that digits of the sensed value (and not other digits illustrating noisy information such as temperature, time and date) are objects of interest, together with a post-processing step to reorder objects and prepare the final response. DNN-based methods have shown remarkable results, but to get the most out of them, we need a well-annotated training dataset, which is not available for our problem.
	We design a better workflow and a data generation algorithm to automatically synthesize thousands of training images with well-aligned annotations. 
	Our data generation workflow is described in detail in \S\ref{sec_data_generation}.

	We use mobile smartphones to capture images from medical devices and to extract useful information. While inference on the mobile device provides low latency and high availability for users, it usually suffers from low accuracy due to resource constraints. On the other hand, inference on the cloud provides accuracy and performability, but high latency and unavailability in the case of poor network connection are its big weaknesses. 
	As depicted in Fig.\,\ref{fig:mobile_cloud_architecture}, our proposed hybrid mobile-cloud serving architecture takes advantage of both mobile and cloud computing. Additionally, we designed a specialized ensemble model to further improve our desired performance metrics.

\subsection{Deep Learning Models}

	We reduce the problem of image-based reading from screen of medical devices to an object detection and post-processing task.
	Each digit and decimal point belonging to the sensed value occupies a region of interest (RoI) and has a corresponding class (Fig.\,\ref{fig:image_based_method}-1). 
	In post-processing step, we convert a group of objects to a meaningful string (Fig.\,\ref{fig:image_based_method}-2). 
	For the object detection part, we design and train two convolutional models (CNNs) based on single-shot detector (SSD) architecture \cite{liu2016ssd}. 
	
	In general, we have powerful resources on cloud, but limitation at the mobile. Thus, we consider two proportionate backbones for our SSD architectures. One is highly optimized for smartphones and has fewer parameters; hence, lower accuracy. Another one is more accurate and consumes more resources so that can not be deployed on mobile.
	We prefer SSD-based networks for both sides, because they achieve better end-to-end latency \cite{Huang2017b}.
	
\subsubsection{Cloud-Side Model}

	We employ ResNet-50 \cite{He2016} as the backbone of our SSD model on the cloud-side. It contains 16 convolutional blocks with shortcuts and one fully connected layer. It has more than $25M$ parameters, and requires $~4$ billion multiplication-accumulation operations (MACs) per sample. 
	We remove its last few layers including the classification head to use the remainder as a feature extractor backbone for the SSD architecture. 
	Since the size of RoIs in input images varies, we utilize feature pyramid network (FPN) \cite{Lin2017}. It leverages intrinsic pyramid structure of modern CNNs (i.e., ResNet) to generate multi-scale feature maps. FPN can improve accuracy of the model, and is much more compute efficient than feeding an image with different sizes multiple times.

\subsubsection{Mobile-Side Model}

	Similar to the cloud-side, we stacked up a CNN-based classification model as a feature extraction backbone for our mobile-side SSD architecture. However, we employed edge-friendly architectures both for the backbone and the detection head.
	Our mobile-side model is directly inspired by the current state-of-the-art detection architecture for mobile devices, the MobileDets \cite{xiong2021mobiledets}. 
	The reason of using depth-wise separable convolutional layers instead of regular convolutional layers was their fewer parameters and MACs, while hoping these metrics directly relate to less train and inference time. 
	However, recent studies \cite{Sze2020, Stamoulis2020} have shown network parameters and MACs may not be good proxies to model inference throughput and latency as they are not the only factors. Therefore, MobileDets expands the neural architecture search space by adding regular convolutions as well. 
	After multiple experiments on different backbone architectures, we found the architecture proposed by \cite{xiong2021mobiledets} for Mobile CPUs is the most appropriate for our mobile-side model. Since mobile devices are resource-constrained, we do not leverage FPN.

\subsection{Ensemble Algorithm}

As illustrated in Fig.\,\ref{fig:mobile_cloud_architecture}, we design and deploy a light-weight engine at the mobile along with a more compute-intensive and accurate one on the cloud. We integrate predictions of our two deep learning models through our ensemble algorithm (Algorithm\,\ref{algo_ensemble_learning}). 
Having regions of interest (RoIs), labels and confidence scores of the objects detected by both models, our ensemble algorithm first finds corresponding RoIs among mobile and cloud predictions (line\,\ref{lst_find_corr_rois}). 
They must have identical labels (e.g., class of digit '5') and their distance should not be more than a tolerable amount $\epsilon$. 
The intuition behind is, the RoI coordinates predicted by different object detection models may not exactly match with each other. 
$\epsilon$ controls maximum distance between two corresponding predictions from two different models. There may exist more than one object with same labels. So if we do not check that, it will lead to misrecognition of RoIs. However, strict comparison is a bad idea because two quite different deep learning models may have small differences in RoI refinement after training. Hence, we add a tolerable distance to make it more suitable.

\begin{algorithm}[b!]
\caption{Ensemble Learning Algorithm}
\label{algo_ensemble_learning}
\begin{small}
\begin{algorithmic}[1]
\REQUIRE
{\small
\encircle{1}~maximum number of RoIs ($N$). 
\encircle{2}~RoIs matrices ($R$), and their corresponding vectors of \encircle{3}~confidence ($\rho$) and \encircle{4}~labels ($L$).
~\encircle{5}~confidence thresholds ($T$). \encircle{6}~tolerable distance ($\epsilon$).
\newline $\triangleright$ Each $R{^K_l}$ comprises four member points: 
left~($x_{min}$), right~($x_{max}$), up~($y_{min}$), down~($y_{max}$).
\newline $\triangleright$ Superscripts $C$, $M$ and $E$ stand for \textit{Cloud}, \textit{Mobile (Edge)} and \textit{Ensemble}, respectively.
}
\ENSURE 
{\small A string equivalent to final prediction (e.g., "10.6").}
\vspace{.5mm} \hrule \vspace{1mm}
\STATE {$R^E \gets \emptyset$,~ $\rho^E \gets \emptyset$,~ $L^E \gets \emptyset$}
\FOR {$i \gets 1$ \TO $N$} 
\STATE {
  Find $j$ such that:
	\newline ~~-$R{^C_i}[x_{min}] \approx R{^M_j}[x_{min}] \pm \epsilon$ \AND
	\newline ~~-$R{^C_i}[x_{max}] \approx R{^M_j}[x_{max}] \pm \epsilon$ \AND
	\newline ~~-$L{^C_i} = L{^M_j}$
\label{lst_find_corr_rois}}
\IF {$j \ne \O$}
  \STATE {$L^E \gets L{^C_i} \cup L^E$,~  $R^E \gets R{^M_j}\cup R^E$}
  \IF {$\rho{^C_i} \geq T{^C_i}$ \AND $\rho{^M_j} \geq T{^M_i}$}
  	\STATE {$\rho^E \gets(\rho{^M_j}+ \rho{^C_i})\cup\rho^E$ \label{lst_add_scores}}
  \ELSE
  	\STATE {$\rho^E \gets\max(\rho{^M_j}, \rho{^C_i})\cup\rho^E$ \label{lst_max_scores}}
  \ENDIF
  \STATE {$R^M\gets R^M\text{\textminus}R^M_j$,~$\rho^M\gets \rho^M\text{\textminus}\rho^M_j$,~$L^M \gets L^M\text{\textminus}L^M_j$}
\ELSE
  \STATE {
  	$L{^E} \gets L{^C_i}\cup L^E$,~ 
  	$\rho^E \gets \rho{^C_i}\cup \rho^E$,~ 
  	$R^E \gets R{^C_i}\cup R^E$
  	\label{lst_attach_cloud}
  }
\ENDIF
\ENDFOR
\STATE {$R^E \gets R^M \cup R^E$,~$\rho^E \gets \rho^M \cup \rho^E$,~$L^E \gets L^M \cup L^E$ \label{lst_attach_mobile}}
\STATE {$I \gets \{ i \in \mathbb{N}~|~i \leq |\rho^E|$,~ $\rho^E_i \geq T^E \}$	\label{lst_universal_thresh}}
\STATE {$L^E \gets  [L^E_{i \in I}~|~\forall (i, j): i < j \Leftrightarrow R^E_i[x_{min}] \leq R^E_j[x_{min}]]$	\label{lst_ensemble_reorder}}
\RETURN {$\|^{n}_{i=1}L^E_i$  {\small \tab[5mm] $\triangleright~$($\|$ concats every element within $L^E$).}	\label{lst_ensemble_return}}
\end{algorithmic}
\end{small}
\end{algorithm}

	After finding, when both scores are higher than specified thresholds, the ensemble algorithm adds them up as the final confidence score (line\,\ref{lst_add_scores}). Otherwise, picks the highest score (line\,\ref{lst_max_scores}). 
	In the case of only one prediction for a particular object on either mobile or cloud, simply adds it to our final results (lines\,\ref{lst_attach_cloud}, \ref{lst_attach_mobile}). 
	The second elimination step happens in (line\,\ref{lst_universal_thresh}), where all remaining objects are compared via a higher universal threshold.
	Eventually, our algorithm reorders the remaining objects by their placement within image (line\,\ref{lst_ensemble_reorder}), and concatenates a sequence of corresponding labels to generate the output string (line\,\ref{lst_ensemble_return}).

\subsection{Post-Processing Algorithm}

	To remove redundant objects more, and to improve accuracy even further, we add an additional post-processing algorithm right before the line\,\ref{lst_ensemble_reorder} in Algorithm\,\ref{algo_ensemble_learning}. We apply a modification of Non-Max Suppression (NMS)~\cite{Neubeck2006} to be compatible with our problem. 
	Here, our objective is to find and remove those RoIs that significantly overlap with each other. It means that there are more than one object of interest in a region while it must not. 
	Leveraging NMS reduces false positive (FP). Our modified NMS is defined in Algorithm\,\ref{algo_postprocess}. It executes iteratively. 
	Having RoIs and their corresponding confidence scores and labels, it first sorts RoIs based on their confidence in a descending order (line\,\ref{lst_pp_sort}). 
	It then removes those objects that their overlap of the occupied area with another object is high enough ($>T_{nms}$), and their confidence is simultaneously lower (lines\,\ref{lst_pp_check_tresh}, \ref{lst_pp_remove_overlapped}). 
	The overlap between two objects is calculated by intersection over union (IoU) of their corresponding RoIs. Although this works well, using a single threshold for all labels introduces a problem in our specific task. 
	The \textit{unit} label, representing the decimal point of sensed numbers, considerably overlaps with other objects, but it must not be removed. This also applies to the classes of '1' and '7'.
  To avoid that, we consider the \textit{unit} label separately in our calculations within NMS procedure (lines\,\ref{lst_pp_unit_start}-\ref{lst_pp_unit_end}).

\begin{algorithm}
\caption{Post-Processing Algorithm}
\label{algo_postprocess}
\begin{small}
\begin{algorithmic}[1]
\REQUIRE
{\small
\encircle{1}~overlap threshold ($T_{nms}$).
\encircle{2}~RoIs matrix ($R$), and its corresponding vectors of \encircle{3}~confidence ($\rho$) and \encircle{4}~labels ($L$).
}
\ENSURE 
{\small reduced RoIs matrix, and its corresponding vectors of confidence and labels.}
\vspace{.5mm} \hrule \vspace{1mm}
\STATE {$R^M \gets \emptyset$,~ $\rho^M \gets \emptyset$,~ $L^M \gets \emptyset$}

\STATE $i \gets j \in \mathbb{N}~|~j\geq |L|, L_j = $'.'	{\small\tab[9mm]$\triangleright~${('.'$\equiv$ decimal point)}
\label{lst_pp_unit_start}}
\IF {$i \neq \O $}
  \STATE {$L^M \gets L_i$,~ $\rho^M \gets \rho_i$,~ $R^M \gets R_i$}
  \STATE {$L \gets L - L_i$,~$\rho \gets \rho - \rho_i$,~$R \gets R-R_i$}
\ENDIF \label{lst_pp_unit_end}
\STATE {$(L, \rho, R) \gets [(L_i, \rho_i, R_i)~|~ \forall (i, j): i<j \Leftrightarrow \rho_i \geq \rho_j ]$	\label{lst_pp_sort}}
\WHILE {$R \neq \emptyset$}
  \STATE {$R^M \gets R_1 \cup R^M$,~ $R \gets R - R_1$}
  \STATE {$L^M \gets L_1 \cup L^M$,~ $L \gets L - L_1$}
  \STATE {$\rho^M \gets \rho_1 \cup \rho^M$,~ $\rho \gets \rho - \rho_1$}
  \FOR {$i \gets 1$ \TO $|R|$}
     \IF {IoU($R_i, R^M_{1}$) $ \geq T_{nms}$	\label{lst_pp_check_tresh}}
        \STATE {$R \gets R - R_i$,~ $L \gets L - L_i$	\label{lst_pp_remove_overlapped}}
     \ENDIF
  \ENDFOR
\ENDWHILE
\RETURN $R^M \cup R$,~ $L^M \cup L$,~ $\rho^M \cup \rho$
\end{algorithmic}
\end{small}
\end{algorithm}
	
  \subsection{Bandwidth Optimization}
	Our target usecase is in underdeveloped countries, thus more often than not, some users may be located in areas or situations that do not have access to the internet or their connection is poor, e.g., due to limited available bandwidth or network congestion problems. 
	Therefore, we prefer providing the highest availability at the cost of less accurate response, to increase performability, in general.
	One can still get the mobile answer in zero-connection situations. Nevertheless, user will experience accuracy degradation.
	For poor network conditions, we present a simple image compression technique which reduces the bandwidth usage when sending captured images (Algorithm\,\ref{algo_compression}).
	We down-scale the image, transform it from RGB to HSV colorspace, and then only send the value ($V$) channel of the image to the cloud. On the cloud, $H$ and $S$ are filled with predefined constants and are up-scaled to the original dimension before performing inference.
	
\begin{algorithm}[h]
\caption{Simple Lossy Compression}
\label{algo_compression}
\begin{small}
\begin{algorithmic}[1]
\REQUIRE
{\small
\encircle{1}~$Img$
\encircle{2}~output size ($\mathcal{H}\nu,\mathcal{W}_\nu$) 
\encircle{3}~filling constant ($\mathcal{K}$)
}
\ENSURE 
{\small
At mobile : $Img_{comp.}$ \quad At cloud : $Img_{decomp.}$
}
\vspace{.5mm} \hrule \vspace{1mm}
\end{algorithmic}
\textbf{At mobile}
\begin{algorithmic}[1]
\STATE $Img_{comp.}\gets$ Resize $Img$ to ($3\times\mathcal{H}\nu\times\mathcal{W}\nu$).
\STATE Transform $Img_{comp.}$ from $RGB$ colorspace to $HSV$.
\STATE $Img_{comp.}\gets$ Drop channels $H$ (Hue) and $S$ (Saturation).
\RETURN $Img_{comp.}$
\end{algorithmic}
\textbf{At cloud}
\begin{algorithmic}[1]
\setalglineno{5}
\STATE $Img_{decomp.} \gets $ Add channels $H$ and $S$, and fill pixels with $\mathcal{K}$.
\STATE Resize $Img_{decomp.}$ to original size of $Img$.
\RETURN $Img_{decomp.}$
\end{algorithmic}
\end{small}
\end{algorithm}

\subsection{Complexity of Proposed Algorithm}

	For every detected object we sweep the other objects. Similarly, this is done again in the post-processing algorithm, for fewer objects. The time complexity of our algorithm, therefore, is $O(N^2{+}{N'}^2)$ where $N$ is the number of predicted RoIs and ${N'}$ is the number of objects to post-process. Since $N'{<}N$, the complexity can be written as $O(N^2)$. We use one-dimensional arrays to keep RoI boxes and confidence and label vectors, that their size depend on $N$. Consequently, the space complexity of the algorithm is $O(N)$.

\section{Data Generation}
\label{sec_data_generation}

	{Data scarcity} in image-based reading is a major reason for poor results in the previous works.
	The superpower of deep learning models (e.g., CNNs) is their ability to learn representations directly from data.
	Deep learning algorithms depend on either big volume of annotated data to be trained in a supervised manner or a model pre-trained on a related task; none are available for our task. 
  
\begin{figure}[!t]
  \centerline{\includegraphics[width=\columnwidth]{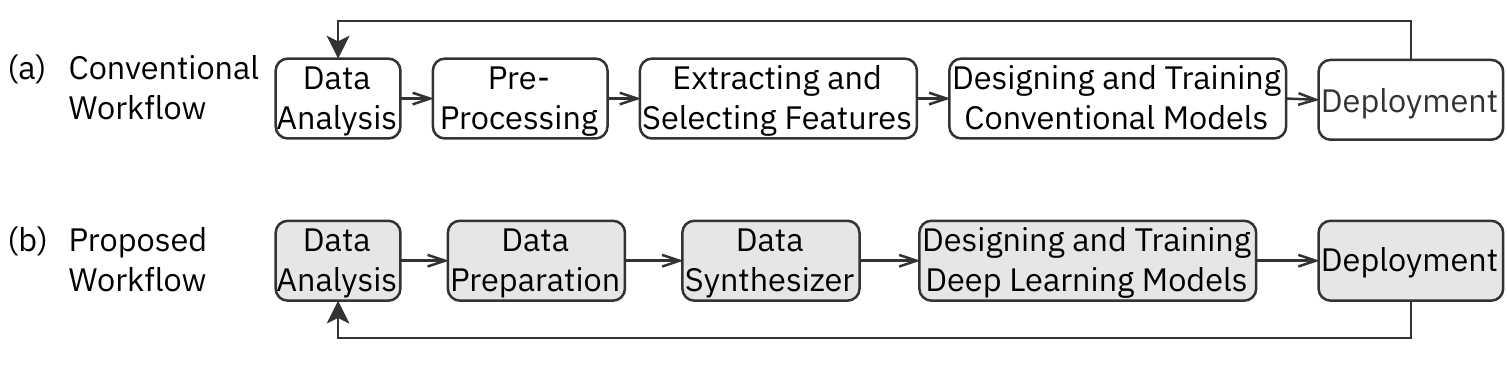}}
  \vskip -0.15in
  \caption{(a)~conventional computer vision workflow. (b)~proposed workflow. Note that although some blocks are the same in both, their internal procedure might be completely different.}
  \label{fig:cv_workflows}
  \end{figure}

	Related work commonly use conventional computer vision workflow which starts with data analysis and subsequently continues with data cleaning, pre-processing, feature extraction, feature selection and finally designing and deploying a suitable model (Fig.\,\ref{fig:cv_workflows}(a)).  
	In contrast, we present a different workflow (Fig.\,\ref{fig:cv_workflows}(b)) which enables us to utilize deep learning models, outperforming previous algorithms by a large margin.
	
\subsection{Data Preparation}

	After analyzing the images captured from displays of various glucometers, we discovered a number of common differentiating properties, summarized in Table\,\ref{table_medical_device_properties}. 
	We learned that a lot of problems stem from the environmental conditions. For example, as depicted in Fig.\,\ref{fig:example_images}, some images have been rotated (a, b, d, f, g), taken from different viewpoints (a, b, d, e, f), have poor contrast between the sensed quantity and its background (a, e, f), and have flashlight reflection (h), distorted and/or blurred (e, h). 
	Some of the patterns from Table\,\ref{table_medical_device_properties} are observable as well. For example, they may contain charts (d), have backlit (c, d, g), irregular screen shapes (g, h). 
	And almost all of them are different in background color, display font style and type, etc. 
	
	This diversity in medical devices together with environmental variations introduce many difficulties for previous methods as they are based on feature engineering.	
	Instead, we utilize such properties to synthesize a large training dataset with annotation. 
	We gather 100 distinct images of medical devices and 150 open-source fonts (including seven-segment, dot-matrix, LCD, and LED styles) through the Internet. 
	We then manually transform all images to obtain images with standard point of view. 
	We also determine ${\thicksim}20$ different point coordinates per image corresponding to display screen corners and different items that can be on it (e.g., sensed value, measurement unit, date, etc.). We eventually feed all these preliminary information into our \textit{Data Synthesizer} algorithm which can synthesize almost infinite distinct images of medical device with different sensed values.

\begin{table}[b]
	\vskip -0.2in
	\caption{Common differentiating properties and troublemakers in image-based readout from medical devices.}
	\label{table_medical_device_properties}
	\setlength{\tabcolsep}{3pt}
	\begin{tabular}{rp{1.4cm}p{6.4cm}}
		\hline
		\#	& Property	& Pattern									\\
		\hline
		1	& Display	& Different aspect ratios					\\
		2	&			& Different screen shapes					\\
		3	&			& LCD, LED, 7-segment, dot-matrix			\\
		4	&			& With or W/O backlit						\\
		5	&			& Regular or irregular screens				\\
		6	&			& Rounded or sharp corners					\\
		7	& $\mathsf{Value}$		& Different measurement units ($mg/dL$, $mmol/L$) \\
		8	& (Label)	& Usually, the biggest item on display		\\
		9	&			& No specific position on different displays\\
		10	& Additional {$Items$} & Include date, time, various signs and symbols, measurement unit, diagrams, etc. \\
		11	&			& No specific position on different displays\\
		12	&			& In some cases, fonts, styles, and colors differ from {$\mathsf{Value}$}	\\
		\hline
		\end{tabular}
	\vskip -0.3in
\end{table}

\subsection{Data Synthesizer}

	Our \textit{Data Synthesizer} (Algorithm\,\ref{algo_synthesizer}) receives the data prepared from the last step (line\,\ref{lst_syn_proc}), namely a number of transformed images ($\mathcal{I}mages$) along with their corresponding point coordinates ($\mathcal{A}$) a set of fonts ($\mathcal{F}$), 
  a degree-of-freedom ($\mathcal{D}$) for every item on the display and their properties,
  and maximum number of images to synthesize ($\mathcal{N}$). 
	It then generates new images (line\,\ref{lst_syn_sample}) and strict annotations (line\,\ref{lst_syn_gen}) following Algorithm\,\ref{algo_synthesizer}.
	Producing thousands of images from a small sample size, particularly to train deep learning models, is an important procedure, because any subtle noise or shift in distribution of training data leads to the overfitting issue: our models may reach to a reasonable accuracy on our synthesized training data, but perform poorly on real-world test data.
	Therefore, all artificially generated images must be quite similar to real images.
	On the other hand, to train the deep learning models, our training set should contain enough distinctive features. 
	Also, our algorithm has to generate well-aligned annotations.
\begin{algorithm}
\caption{Data Synthesizer}
\label{algo_synthesizer}
\begin{small}
\begin{algorithmic}[1] 
  \REQUIRE
  {\small
  \encircle{1}~source images~$\mathcal{I}mages$,
  \encircle{2}~corresponding coordinates~$\mathcal{A}$, 
  \encircle{3}~set of fonts~$\mathcal{F}$,
  \encircle{4}~degrees-of-freedom~$\mathcal{D}$, and
  \encircle{5}~maximum number of samples to generate~$\mathcal{N}$.
  }
  \ENSURE 
  {\small A set of $\mathcal{N}$ tuples each containing a synthesized image, its label ($\mathsf{Value}$), and corresponding RoIs ($R^{\mathsf{Value}}$,$R^{Items}$).}
  \vspace{.5mm} \hrule \vspace{1mm}
\PROCEDURE{Synthesizer}{$\mathcal{I}mages^{\mathcal{S}et},\mathcal{A},\mathcal{D}, \mathcal{F}, \mathcal{N}$} \label{lst_syn_proc}
 \FOR {$\iota \gets 1$ \TO $\mathcal{N}$}
   \REPEAT
     \STATE $(Img,R^{Screen})\gets\mathrm{Sample}_{Rand}(~(\mathcal{I}mages^{\mathcal{S}et}, \mathcal{A})~)$ \label{lst_syn_sample}
     \STATE $(\mathsf{Value},Items)\gets\mathrm{Generate}_{Rand}(Img,\mathcal{A}_{Img},\mathcal{D}_{Img},\mathcal{F})$ \label{lst_syn_gen}
     \STATE $\forall$\tab[.5mm]item\tab[0.5mm]$j\in Items~\forall$\tab[.5mm]object\tab[.5mm]$k\in$\tab[.5mm]item\tab[.5mm]$j:\mathrm{Calculate}(R^{Items}_{j,k})$.
     \STATE $\forall$~object~$i\in \mathsf{Value}:\mathrm{Calculate}(R^{\mathsf{Value}}_i)$.
   \UNTIL {$R^{\mathsf{Value}}_i \cap R^{Items}_{j,k} = \emptyset$ \AND \newline \tab[5mm] IoU($R^{Items}_{i}$, $R^{Items}_{j}$) $<\epsilon$,~ $i\neq j$	\label{lst_syn_val_1}}
   \REPEAT
     \STATE $Img \gets \mathrm{Transform}(Img,~\mathcal{S}et)$ \label{lst_syn_transform}
     \STATE $\forall R_i\in R^{\mathsf{Value}}:\mathrm{Calculate}(R_i)$\tab[9mm]{\small $\triangleright$~(Recalculates~RoIs.)}
   \UNTIL {$\forall R_i \in R^{\mathsf{Value}}:~R_i \cap (\mathbb{U}-R^{Screen})=\emptyset$}
   \STATE \textbf{store} $\{Img,R^{\mathsf{Value}},\mathsf{Value},R^{Items}\}$\tab[3mm]$\triangleright$~$\mathsf{Value}\equiv$Label $L_{Img}$
 \ENDFOR
\ENDPROCEDURE
\end{algorithmic}
\end{small}
\end{algorithm}

We use $\mathcal{D}$ to make sure the synthesized images and their corresponding annotations are both similar to real captured images and follow the patterns described in Table\,\ref{table_medical_device_properties}.
We introduce controlled stochasticity to our synthesizing procedure. Randomness is a key enabler to increase mutual distinction between training samples, to prevent overfitting and improve eventual accuracy. 
	Additionally, our algorithm randomly selects or generates different font styles, sizes, colors, background lights, time, date, and different indicators (e.g., temperature or blood drop symbol) (line\,\ref{lst_syn_gen}). The font styles, font sizes, postions, and colors of additional information are selected from a range randomly as well.
	It validates the placement of each item afterwards (line\,\ref{lst_syn_val_1}).
	There must not be any overlap between the sensed $\mathsf{Value}$ and other additional information on screen ($Items$) such as time, date, temperature, etc. Also, there should not be any significant ($>\epsilon$) intersection between two different $Items$.

\begin{figure*}[t!]
  \centering
  \subfigure[]
  {
      \includegraphics[width=0.105\textwidth]{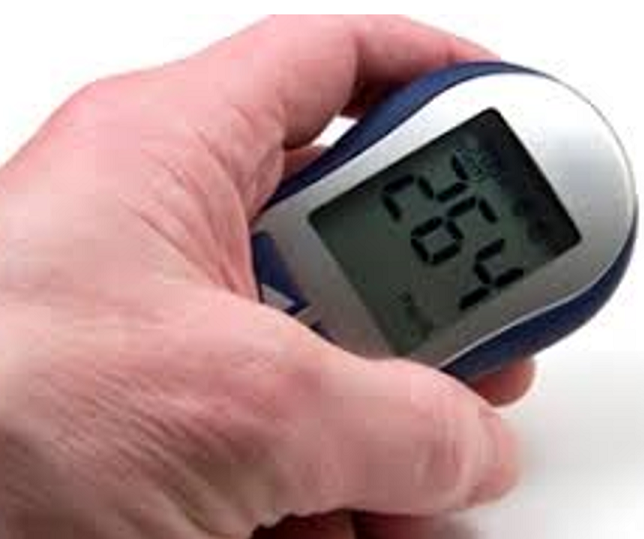}
      \label{fig:synthesized_samples_7a}
  }
  \subfigure[]
  {
      \includegraphics[width=0.105\textwidth]{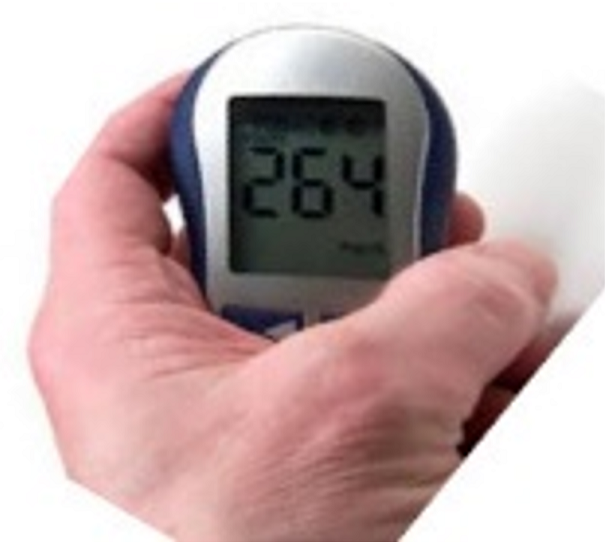}
      \label{fig:synthesized_samples_7b}
  }
  \subfigure[]
  {
      \includegraphics[width=0.105\textwidth]{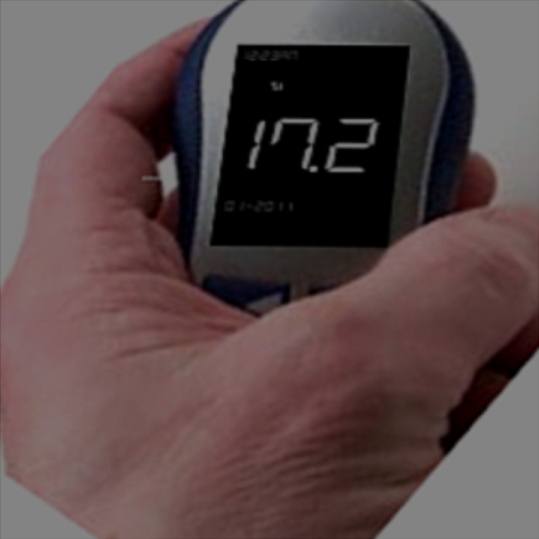}
      \label{fig:synthesized_samples_7c}
  }
  \subfigure[]
  {
      \includegraphics[width=0.105\textwidth]{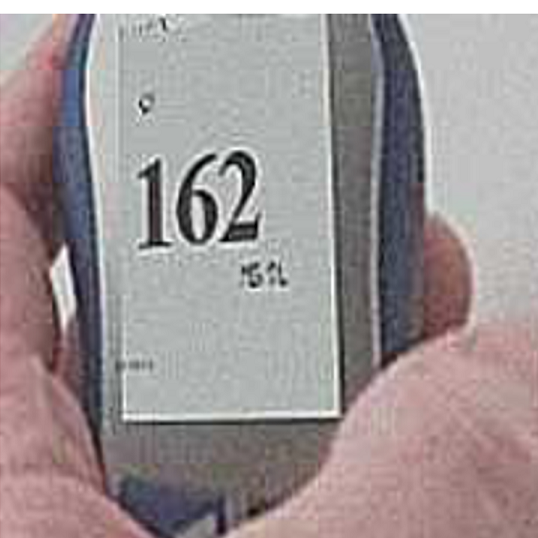}
      \label{fig:synthesized_samples_7d}
  }
  \subfigure[]
  {
      \includegraphics[width=0.105\textwidth]{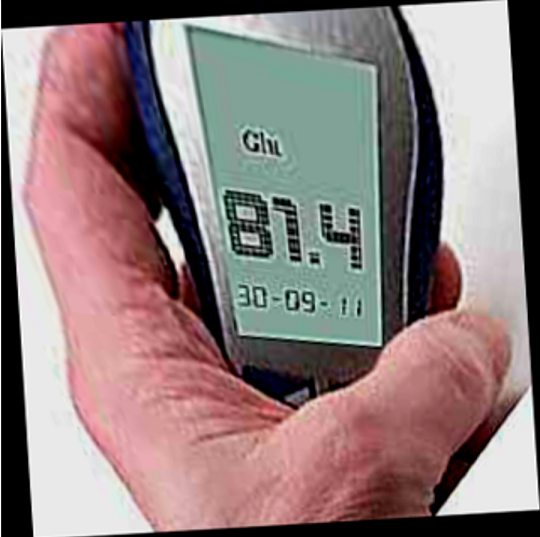}
      \label{fig:synthesized_samples_7e}
  }
  \subfigure[]
  {
      \includegraphics[width=0.105\textwidth]{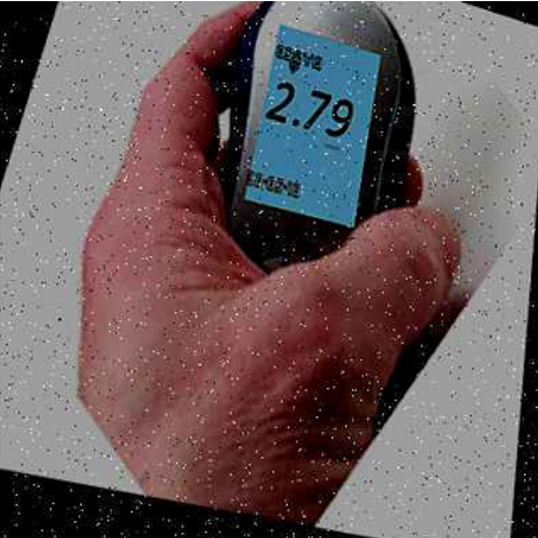}
      \label{fig:synthesized_samples_7f}
  }
  \subfigure[]
  {
      \includegraphics[width=0.105\textwidth]{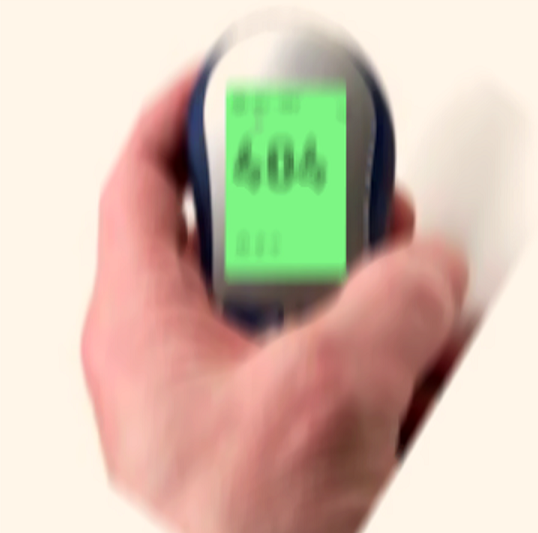}
      \label{fig:synthesized_samples_7h}
  }
  \subfigure[]
  {
      \includegraphics[width=0.105\textwidth]{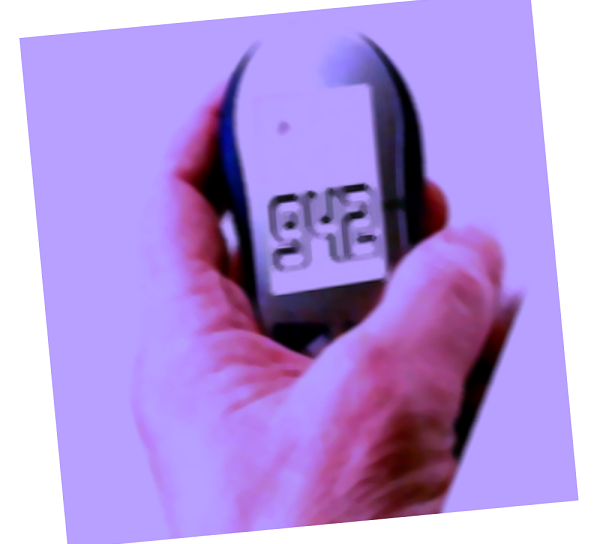}
      \label{fig:synthesized_samples_7i}
  }
\vskip -0.15in
\caption{A few examples of the images synthesized and transformed by Algorithm\,\ref{algo_synthesizer}. (a) is the original image, and (b) is its manually transformed image. Note that for better understanding, only images generated from one source image are shown.}
\label{fig:synthesized_samples}
\vskip -0.15in
\end{figure*}

	Consequently, it calls a transformation procedure (line\,\ref{lst_syn_transform}) that modifies each generated image together with its annotation.
	Our transformation procedure consists of two major sections: \textit{geometrical} and \textit{visual}.
	In the geometrical part, we apply scaling, cropping, rotation, shearing, perspective transformation, and translation. In the visual part, we modify color, contrast, lightness, and sharpness. We insert noise, and arbitrarily drop some pixels to simulate light reflections or scratches in display screen. The range and likelihood of each transformation depend on the generation set ($\mathcal{S}et$). In general, we designate a broader range and higher likelihood in the training set rather than validation set. The intuition behind is, we aim to generalize in the training time, while in validation, a set of data much more similar to real-world images is needed.
	Note that we never apply our \textit{Data Synthesizer} on test set. In addition, to prevent any meaningful leakage, we make sure there is no similarity in medical devices in test set and training/validation set.
	After applying transformations, $\mathsf{Value}$ must still be within $Screen$ RoI ($R^{Screen}$). Otherwise, we apply transformation again until the condition is met.
	A few synthesized samples, generated using only one image from $\mathcal{I}mages$, is depicted in Fig.\,\ref{fig:synthesized_samples}.

  $\mathsf{Value}$ in synthesized images, ranges from 0 to 1000 and follows \textit{discrete pseudo iid} distribution. Half of the numbers are integers. 35\% are single decimal (e.g., 12.5), and remainders are double decimal (e.g., 1.25). The generated measurement unit must be compatible with its synthetic $\mathsf{Value}$.
	Backlit and display background colors are not uniformly random, rather our algorithm selects colors that are more likely to appear in real-world devices with higher probability, but there is a probability to generate completely new colors in order to make the models robust against new devices or outliers. 
	Each item~$\in Items$ may randomly appear or disappear on $Screen$. 
	The $\mathcal{I}mages$ are split into train ($\mathcal{I}mages^{train}$) and validation ($\mathcal{I}mages^{val}$) parts, in 90 and 10 shares, respectively. Each image comprises a different medical device.
  \section{Evaluation}
  \label{sec_evaluation}
  
  \subsection{Experimental Setup}
  
    \noindent \textbf{Training Dataset.}
    We generated $1M$ distinct well-annotated training samples using our \textit{Data Synthesizer}. 
    The image size directly affects training throughput, inference latency, size of detection models, and the space needed to store the dataset. Therefore, we down-scale each image to $3\times320^2$ and save in JPEG format. Note that the input sizes of our CNN models are different.  
    It occupies $45GB$ disk space and takes ${\thicksim}11$ hours to generate and store $1M$ samples.
    \newline \noindent \textbf{Test Dataset.}
    We collected 300 images directly captured by our clients from various glucometer devices without any modifications except down-sizing to the desired scale. As Fig.\,\ref{fig:datasets} depicts, the class of '1' has the most frequency of occurrence, and the class of '.' has the least among others. 
    Additionally, we evaluate our method on another publicly available dataset from CameraLab from University of Oxford (referred to as Oxford for simplicity from now on) \cite{finnegan2019automated}.
    \newline \noindent \textbf{Training.}
    We trained both our mobile and cloud neural models on a single server. We use an in-house GPU server (Table\,\ref{table_cloud_spec}) to train our models.
    We designed and trained our models via TensorFlow framework. We reduce the time and resources needed for training the models by employing transfer learning from pre-trained weights on COCO dataset \cite{Lin2014}. It helps our models extract better low-level features. We trained each model for 500-600k epochs (${\thicksim}3.5$ days), and applied simple augmentations during training to prevent overfitting.
    \newline \noindent \textbf{Inference Setup.}
    For the proof of concept, we use the same server as the cloud side.
    For the mobile side, we use Galaxy Tab A (2019), a mediocre tablet (Table\,\ref{table_edge_spec}). We leveraged TensorFlow Lite for Android to convert, optimize, measure the performance of our mobile model.
    During the inference, input images get resized to $3\times416^2$ and $3\times350^2$ for the cloud and mobile models, respectively.
    
  \subsection{Accuracy Metric}
    We report the accuracy of our algorithm as formulated in \eqref{eq_acc_metric_2}. It is more illustrative in comparison with the previously reported formulas, namely Precision, Recall and F1-score for classification and localization. That said, it is the strictest as well, and is necessary due to medical nature of the values being read.
    
    Given a prediction $\hat{Y}$ and its ground-truth $\Lambda$. Let:
  \begin{equation*}
  \label{eq_acc_metric}
  \begin{split}
  \hat{Y} := \hat{y_1}\hat{y_2}\cdots\hat{y_m} \\
  \Lambda := \lambda_1\lambda_2\cdots\lambda_n \\
  \end{split}
  \end{equation*}
  $\hat{Y}$ is then considered as a \textit{correct} prediction if:
  \begin{equation}
  \label{eq_acc_metric_2}
  m=n \text{\quad and \quad} \forall i: \hat{y_i}=\lambda_i,~i\in\mathbb{N},~i\leq n
  \end{equation}
    A prediction is correct if \encircle{1}~all ground-truth RoIs are correctly detected, \encircle{2}~the labels are correctly predicted, \encircle{3}~objects are in the same order as they are in the ground-truth, and \encircle{4}~no additional object (e.g., from $Items$) is detected. 
    For instance, assuming a ground-truth label \textit{11.9}	(Fig.\,\ref{fig:image_based_method}), none of 119, 1.19, 111, 1149, 1109, 11, 19, 1.9, etc. are correct predictions, which makes it more difficult than Precision or Recall.
  
  \subsection{Accuracy on Our Test Set}
  
  \begin{figure}[t]
  \centerline{\includegraphics[width=\columnwidth]{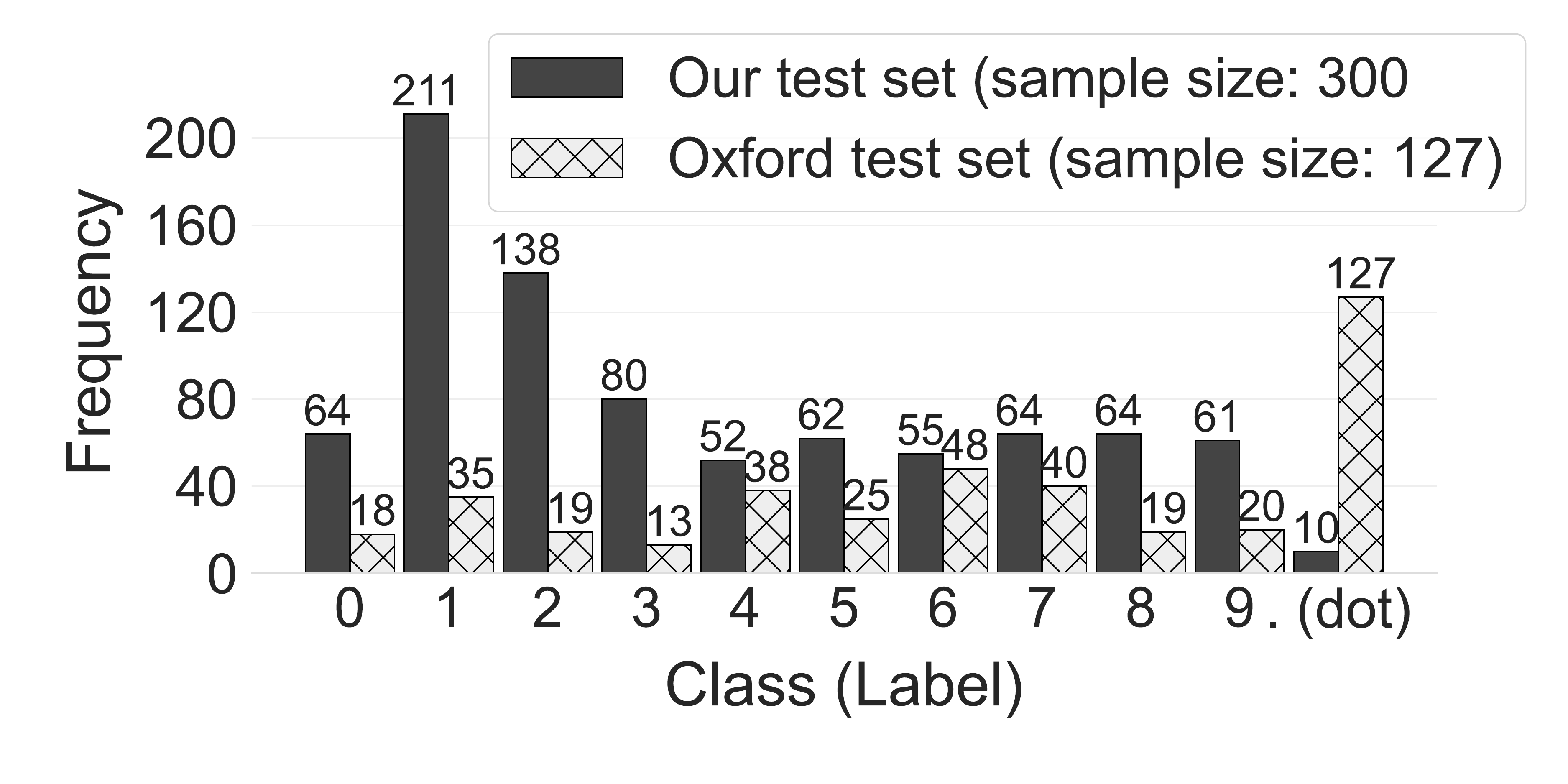}}
  \vskip -0.3in
  \caption{Overview of the test sets used for evaluation. Our dataset contains 300 samples, and Oxford's \cite{finnegan2019automated} contains 127.}
  \label{fig:datasets}
  \vskip -0.15in
  \end{figure}
  
\begin{figure}[t]
  \centerline{\includegraphics[width=\columnwidth]{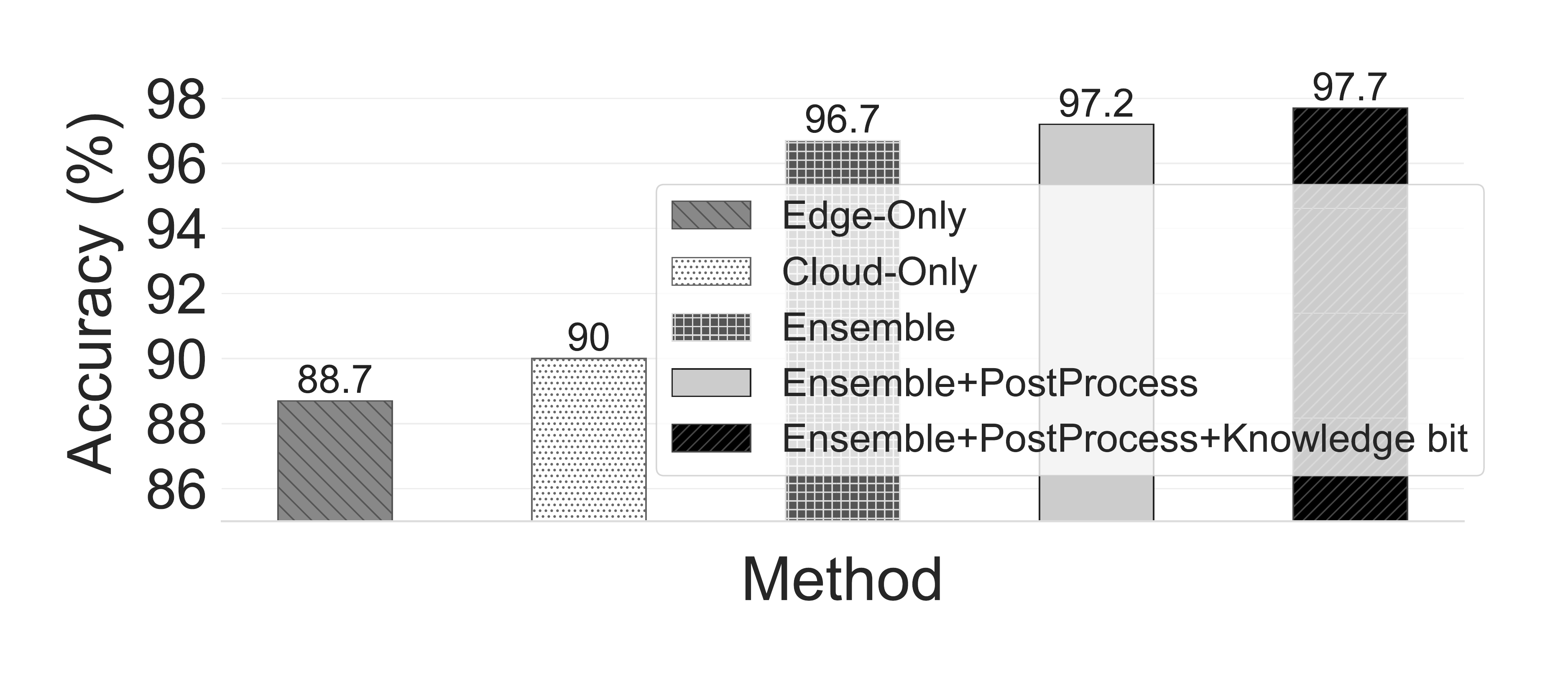}}
  \vskip -0.3in
  \caption{Accuracy on our test set in different settings.}
  \label{fig:accuracy}
  \vskip -0.15in
  \end{figure}

    We first evaluate our algorithms against 300 images directly captured by our clients from various glucometer devices without any modifications except down-sizing to the desired scale. As Fig.\,\ref{fig:datasets} depicts, the class of `1' has the most frequency of occurrence, and the class of `.' has the least among others. 
    While cloud- and mobile-only predictions fluctuate around $89-90\%$ (Fig.\,\ref{fig:accuracy}), our ensemble model improves the accuracy by $7.3$ percentage points. 
    The reason behind is illustrated in Fig.\,\ref{fig:conf_matrix}. In each confusion matrix, every column represents objects in a predicted class while each row represents the objects in its actual class. 
    Making matrices more informative, we added another row and column. The last column represents the objects in a true class that are ignored, and the last row shows the detected objects that do not belong to any classes. 
    To get a better perspective, each matrix is scaled to one in rows. 
    Our cloud model performs well on most classes except `'.' and `1'. 
    On the other side, our mobile model performs poorly on `0', `4' and `9', but predicts `.' and `1' much better than the cloud. 
    The last matrix confirms that our ensemble algorithm perform better than the single models. 
    It also reduces false positives and false negatives. The mispredicted images usually contain some special objects/symbols on their screen. For example, the device shown in Fig.\,\ref{mis_106} displays an arrow on the screen which is pretty similar to `7', and sometimes misleads the models.
    
  \subsubsection{Post-Processing Impact}
\begin{figure*}[t!]
      \centering
      \subfigure[Cloud Model]
      {
          \includegraphics[width=0.3\textwidth]{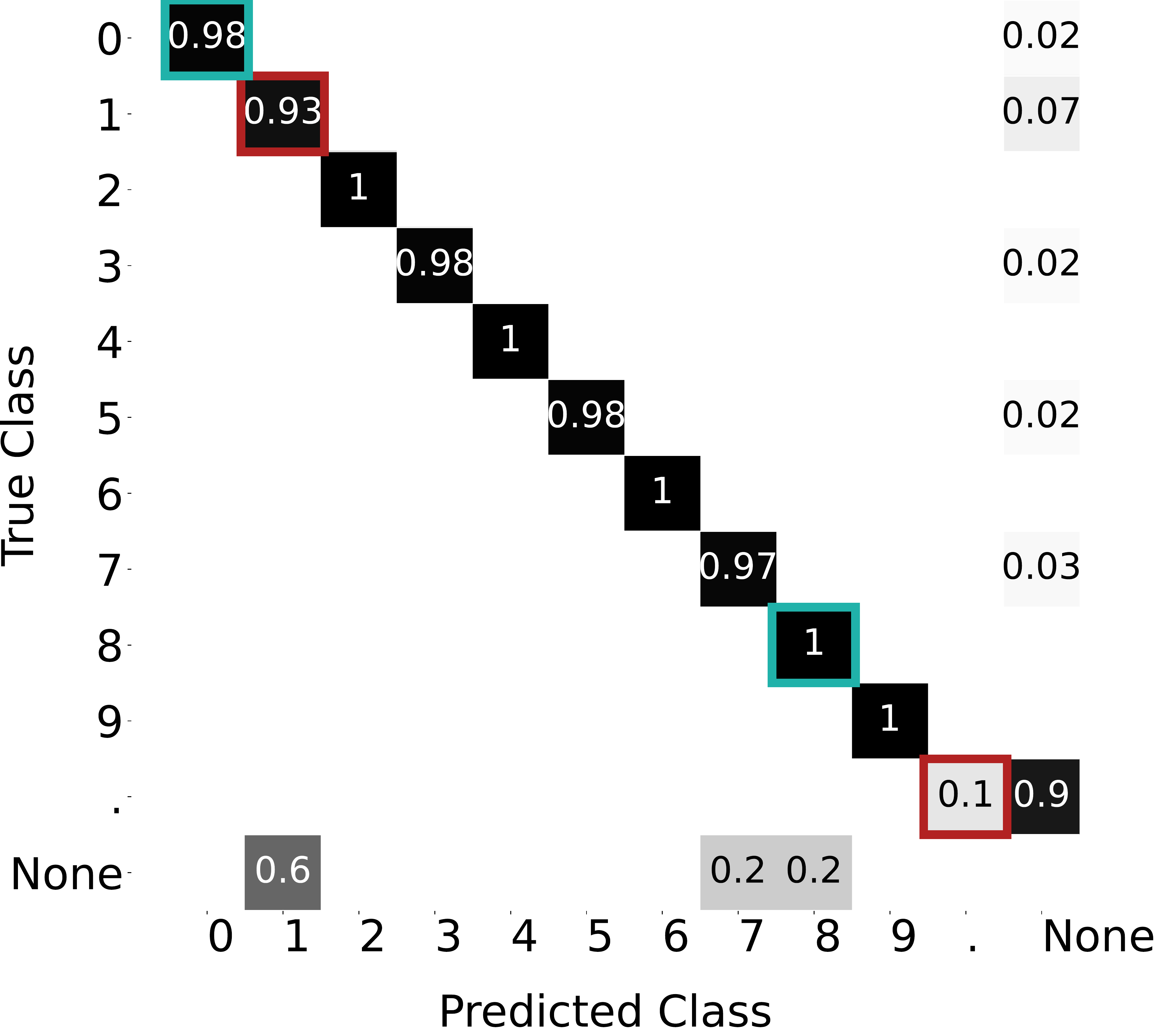}
          \label{fig:confMat_cloud}
      }
      ~
      \subfigure[Mobile Model]
      {
          \includegraphics[width=0.3\textwidth]{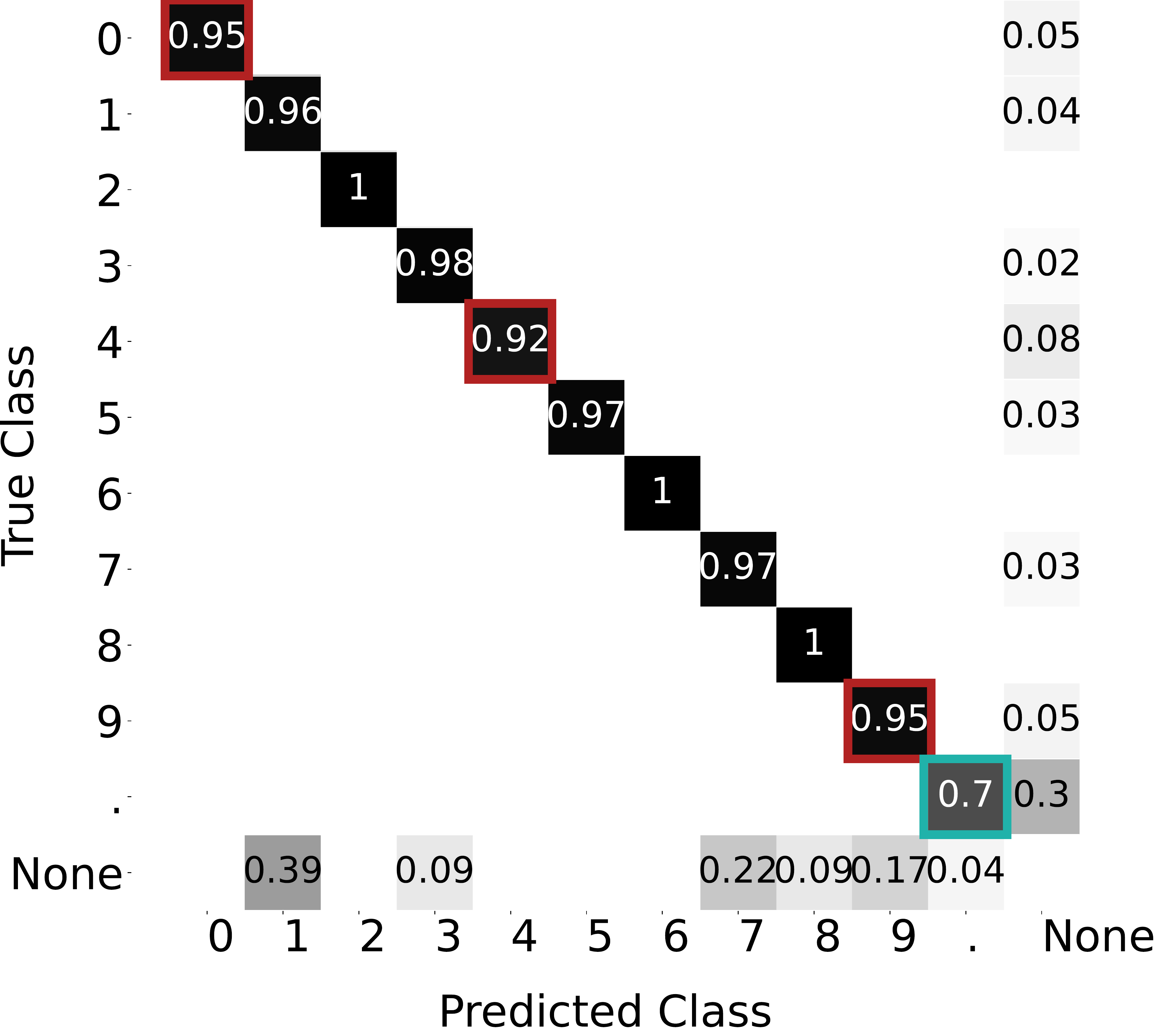}
          \label{fig:confMat_edge}
      }
      ~
      \subfigure[Ensemble Model]
      {
          \includegraphics[width=0.3\textwidth]{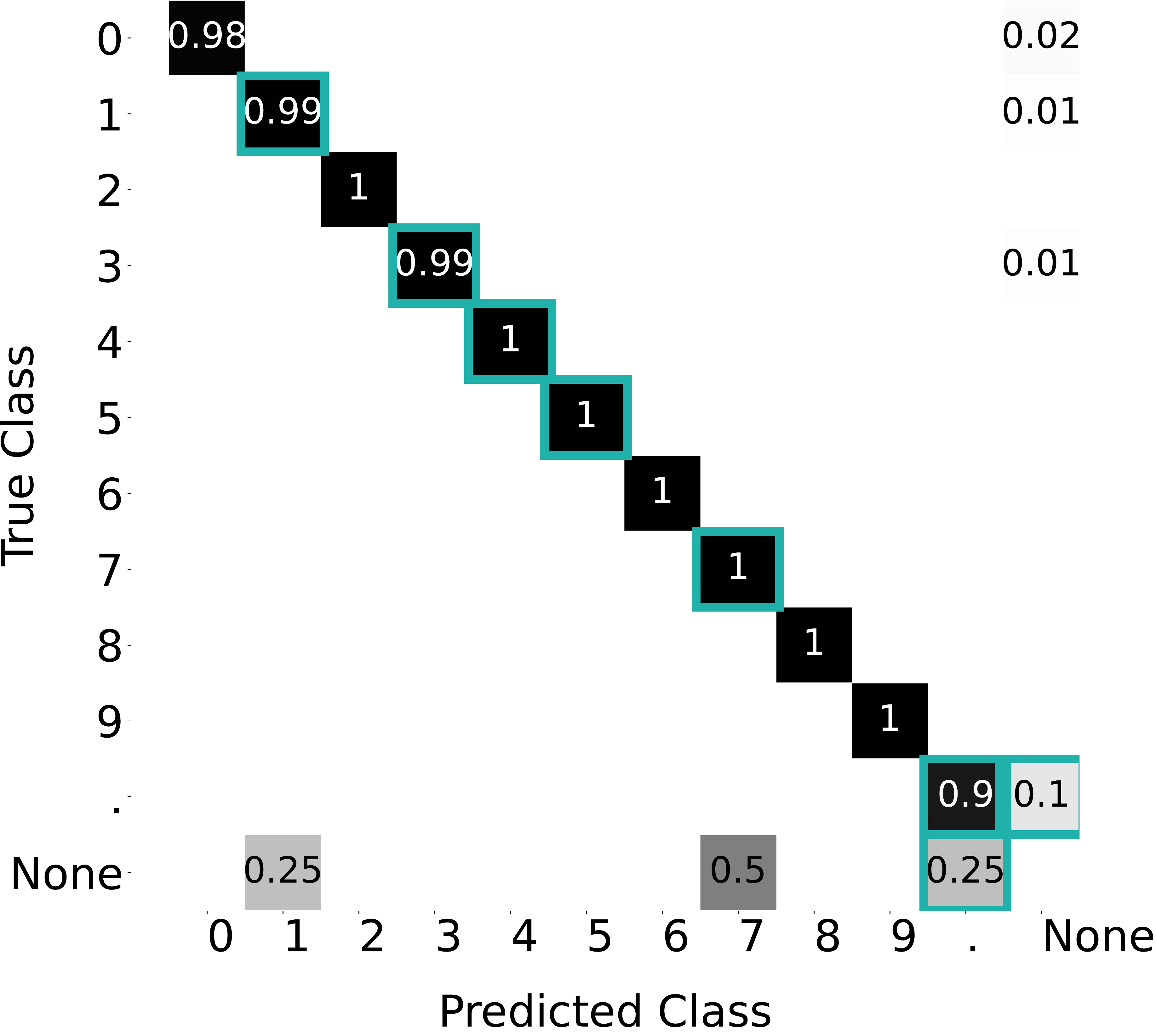}
          \label{fig:confMat_ensemble}
      }
  	\vskip -0.15in
      \caption{Confusion matrix of different models. Matrices are row normalized.}
      \label{fig:conf_matrix}
    \vskip -0.2in
  \end{figure*}

    Our post-processing algorithm shows its positive effect on prediction of test samples. It increases accuracy by $0.5$ up to $1$ on our test set (Fig.\,\ref{fig:accuracy}). The main reason for fewer false positive errors is applying post-processing after the ensemble step (Fig.\,\ref{fig:conf_matrix}).
  
  \subsubsection{Device-Aware Prediction (Knowledge Injection)}
    The least accurate prediction in our proposed method belongs to the class `.' (decimal point) with $90\%$ accuracy. 
    The current state-of-the-art \cite{finnegan2019automated} only detects objects and simply adds a decimal point manually. This is because their dataset contains two devices that both use $mmol/L$ as measurement unit, meaning that all values are floating point. 
    In contrast, our dataset is more general and comprises $mg/DL$ as well. 
    We can improve the accuracy of our algorithm by injecting one bit prior knowledge about the measurement unit of a medical device ($mg/DL$ or $mmol/L$). As shown in Fig.\,\ref{fig:accuracy}, it increases the accuracy by $0.5$. However, considering the displayed measurement unit on the screen as another object of interest can be used in future work.
    
  \subsection{Generalization: Accuracy on Oxford's Dataset}
    
    the Oxford's dataset \cite{finnegan2019automated} gives us the opportunity to evaluate robustness and generalization of our proposed algorithm. 
    We can also have a direct comparison with the current state-of-the-art \cite{finnegan2019automated}. 
    It consists of two different glucometer devices, each evenly divided into training and test sets. The two sets are quite similar (including the image capturing device), except the values on the screens. 
    Each set contains 127 samples, and all of them use mmol/L as the measurement unit on seven-segment displays (Fig.\,\ref{fig:datasets}). 
    The size of each image is ${\thicksim}3{\times}2600{\times}4600$ pixels.
    We first report the accuracy of our models on the test set without learning the training set, to measure how robust our method is. We achieved \textbf{89.8\%} accuracy in this configuration. 
    Next, to get a fair comparison, we fine-tuned our models on the training set.
    In both cases, knowledge injection was not employed. All parameters except $T^E$ were fixed between the evaluations on the two datasets, we used $0.72$ for Oxford's and $0.82$ for ours.
    To prevent Class Imbalance Problem, the Oxford’s training set was combined with a small part of our own training set. After fine-tuning the accuracy improved to \textbf{92.1\%}, while the previous method achieved $51.5\%$ \cite{finnegan2019automated}. 
    As the results suggest, our algorithm outperforms the state-of-the-art both with (+$40.6\%$) and without observing the training set (+$38.3\%$).
    Detailed assessment showed that in most misread images, devices are far from the camera so that in some cases it is difficult to read them correctly. 
    The main reason is lower resolution of our models. Input resolution significantly impacts accuracy of RoI detection \cite{Huang2017b}. 
    In general, reducing size of image by a factor of $4$ decreases accuracy by almost $16\%$ and lowers response time by ${\thicksim}27\%$.
    The input resolution of our models are ${<}1\%$ of the images within Oxford's dataset. Thus, one may be able to achieve even lower error rates by increasing the input size of the neural models, but at the cost of longer training and response time. The memory and storage needed at the mobile-side are also the constraints that must be provisioned.

  \subsection{Availability and Performability}
  \label{subsec:availability_performability}
  
\begin{figure}[t]
  \centerline{\includegraphics[width=\columnwidth]{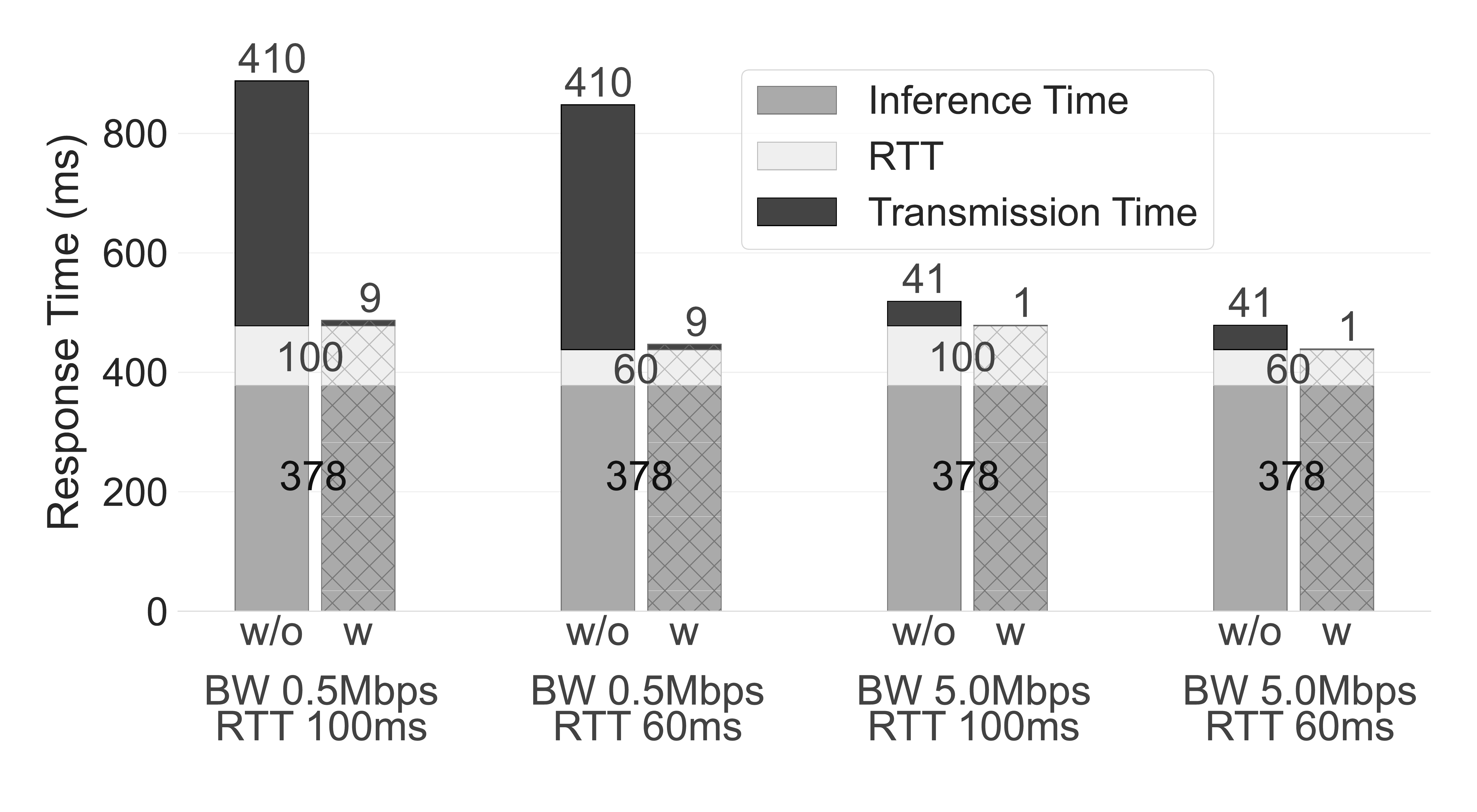}}
  \vskip -0.3in
  \caption{Cloud-side response time breakdown with (w) and without (w/o) applying compression, in two different bandwidth and RTT settings.
  RTT is the round-trip time, and can be measured by the command \textit{ping}. 
  Two RTT values of 100ms, 60ms assumed for illustration.}
  \label{fig:response_breakdown}
  \vskip -0.2in
  \end{figure}

    Our mobile device is able to prepare its prediction in $260ms$ and $360ms$ in GPU-enabled and only-CPU settings, respectively.
    For the situations that there is no internet connection, clients can still rely on the results of their smartphones. 
    They will lose some accuracy, while higher availability can be achieved. 
    When the connection quality is poor, users can enable our simple compression algorithm. It can reduce the bandwidth usage by $\mathbf{45{\times}}$ while resulting $\mathbf{{<}1\%}$ degradation in accuracy, or by $90{\times}$ and ${\thicksim}2\%$ accuracy loss.
    This is achieved by our \textit{Data Synthesizer} algorithm. It forces our deep learning models to learn channel-invariable features. In addition, the ensemble approach inherently improves robustness. 
    The response time breakdown in Fig.\,\ref{fig:response_breakdown} depicts that by applying our compression technique, the transmission delay, which is the major contributor to total delay, collapses.
    Hence, users can still get a reasonable response in ${<}500ms$ with $512Kbps$ available bandwidth and $100ms$ RTT.
  
    To compare related service architectures (Fig.\,\ref{fig:serving_architectures}), we assume a SLA in which \textit{Performability} is described as 
    \textit{reaching $L_k$ accuracy while being able to prepare response within $5s$}. \textit{Availability} can be achieved by responding in ${<}500ms$, as a soft deadline.
    Here, $L_1$ (\dblcheck) and $L_2$ (\snglcheck) stand for ${>}90\%$ and ${>}85\%$ accuracy, respectively. We evaluate each service in three different connection qualities: excellent, poor (e.g., due to congestion or limited bandwidth) and zero (no connectivity). As summarized in Table\,\ref{table_avail_performab}, our model performs better in all three cases. To the best of our knowledge, no comparable Split Computing or Early Exit method currently significantly compresses its intermediate data while losing little to no accuracy (rows 4 and 6).
  
    \begin{table}[b]
      \vskip -0.2in
      \caption{Overview of Availability (Avail.) and Performability (Perf.) in different methods and connections (Conn.).}
      \label{table_avail_performab}
      \setlength{\tabcolsep}{3pt}
      \begin{tabular}{lcc|cc|cc}
        \hline
        \multirow{2}{*}{Service}	& \multicolumn{2}{c|}{Excellent Conn.}	& \multicolumn{2}{c|}{Poor Conn.}	& \multicolumn{2}{c}{Zero Conn.} \\\cline{2-7}
                & Avail. & Perf. & Avail. & Perf. & Avail. & Perf. \\
        \hline
        Mobile-Only		& \dblcheck & \snglcheck & \dblcheck & \snglcheck & \dblcheck & \snglcheck \\
        Cloud-Only		& \dblcheck & \dblcheck$^*$ & $\downdownarrows$ & \dblcheck$^*$ & \xmark & \xmark \\
        Split Comp. 	& \dblcheck & \dblcheck$^*$ & \xmark & \xmark & \xmark & \xmark \\
        Split Comp.$^+$	& \dblcheck & \dblcheck$^*$ & $\downdownarrows$ & \dblcheck$^*$ & \xmark & \xmark \\
        Early Exit		& \dblcheck & \dblcheck$^*$ & \dblcheck & \xmark or\snglcheck & \dblcheck & \xmark or\snglcheck \\
        Early Exit$^+$	& \dblcheck & \dblcheck$^*$ & \dblcheck & \dblcheck$^*$ & \dblcheck & \xmark or\snglcheck \\
        \textbf{Ours}	& \dblcheck[black] & \dblcheck[black] & \dblcheck[black] & \dblcheck[black] & \dblcheck[black] & \snglcheck[black] \\
        \hline
    \multicolumn{7}{p{251pt}}{$^+$\- Assuming intermediate data were compressed by a lossless method.} \\ 
    \multicolumn{7}{p{251pt}}{$\downdownarrows$\- Noticeable degradation.} \\
    \multicolumn{7}{p{251pt}}{$^*$\- Poses extra cost.}
        \end{tabular}
      \vskip -0.3in
    \end{table}

  \section{Conclusion}
  \label{sec_conclusion}
  
      We presented a mobile-cloud automated image-based glucometer reading system broadly applicable to various medical devices; our method uses the camera, the wireless communication, and the computing capabilities of the mobile phone to provide a low-cost alternative to expensive devices available more in developed countries. 
    Our deep learning-based ensemble algorithm together with our mobile-cloud service architecture achieves higher availability and performability compared with mobile-only, cloud-only, split computing, and early exit rival models.
    We proposed a data generation algorithm to address the data scarcity problem, and synthesized one million well-annotated samples.
    Note that the massive varieties in glucometer devices and their screen outputs, has conventionally posed challenges to applicability of existing techniques, but we instead took benefit from them in our data generation techniques to produce more training samples from existing photos, and thus to achieve a more robust model.
    
    Our method is capable of proper readout even in imperfect conditions such as dark ambience, reflections on the screen, blurry and out-of-focus photos. Our ensemble algorithm efficiently combines results obtained from two separate DNN models. Specifically, we take into account the slight shifts in bounding boxes identified by the two models, as well as the special case of `.' symbol that, unlike other symbols, can significantly overlap with other detected objects.

    We evaluated the accuracy of our method on two different real-world test sets, one collected from our users, and one from Camera Lab~\cite{finnegan2019automated}. Our method achieved $97.7\%$ on our test set, and $92.1\%$ on Oxford's, outperforming the current state-of-the-art. Our results showed that our algorithm is robust and generalizes well on other datasets.

\bibliographystyle{IEEEtran}
\bibliography{IEEEabrv, IEEE_TMC.bib}

\appendices

\newpage

\label{sec_appendix}

\section{Mobile Device and Cloud Specifications}

\begin{table}[h]
	\vskip -0.2in
	\caption{Mobile Device Specification}
	\label{table_edge_spec}
  \vskip -0.1in
	\begin{tabular}{p{1cm}p{7cm}}
		\hline
		\multicolumn{2}{c}{\textbf{Samsung Galaxy Tab A 8.0 with S Pen (2019, SM-P205)}} \\
		\hline
		CPU		  & Octa-core big.LITTLE \newline (2${\times}$1.8GHz Cortex-A73 and 6${\times}$1.6GHz Cortex-A53) 	\\
		Chipset	& Exynos 7 Octa 7904 (14 nm) – 64bit\\
		GPU   	& ARM Mali-G71 MP2								  \\
		Memory	& 3GB LPDDR4X									     	\\
		OS		  & Android 10										    \\
		Network	& 2G/3G/4G(LTE) \& Wi-Fi 802.11			\\
		\hline			
	\end{tabular}
\vskip -.5cm
\end{table}

\begin{table}[h]
	\caption{Cloud Server Specifications}
	\label{table_cloud_spec}
  \vskip -0.1in
	\begin{tabular}{p{1cm}p{7cm}}
		\hline
		CPU		& Intel Xeon E5-2630-v3 (x86\_64)		   	\\
				  & Frequency: 2.4GHz (1.2-3.2GHz)			 	\\
				  & Cache: 32K-32K-256K-20480K					 	\\
				  & \#Cores: 24 (double threaded)				 	\\
		GPU 	& NVidia GeForce GTX 1080 Ti				 	  \\
			   	& Total Memory: 11GB							   	  \\
				  & CUDA V9.0.176 , Compute Capacity 6.1	\\
		Memory& 32GB (1600MHz)								   	    \\
		OS		& Gnu/Linux Ubuntu 18.04							  \\
		\hline			
	\end{tabular}
\vskip -0.5cm
\end{table}

\section{Response Time}
Table~\ref{table_response_time} shows total response time on the mobile and cloud in different settings. For mobile inference, Tensorflow benchmarking tool for Android was used, and for the cloud, we measured the time it takes from sending a REST request to the server to get a response. 
The reported response time at the cloud side in TABLE 4 is the summation of inference time, typical transmission time (considering $5$Mbps bandwidth), and typical round trip time (RTT equals $60$ms).

\begin{table}[h]
	\caption{Response Time in Different Settings.}
	\label{table_response_time}
	\vskip -0.1in
	\begin{tabular}{lp{3cm}p{4cm}}
		\hline
		Engine & Platform & Response Time (ms) \\
		\hline
		\multirow{3}{*}{Mobile} & CPU (1 Thread) & $494 \pm 1$ \\
		& CPU (4 Threads) &  $360 \pm 25 $ \\
		& \textbf{GPU (1 CPU Thread)} & $\mathbf{260 \pm 1}$ \\
		\hline			
		\multirow{1}{*}{Cloud} & CPU & $\mathbf{479 \pm 20}$\\
		\hline
	\end{tabular}
\vskip -0.5cm
\end{table}

\section{Bandwidth Optimization}
Table\,\ref{table_bw_opt} illustrates how our compression method affects bandwidth and accuracy with respect to different input resolutions.

\begin{table}[b!]
\caption{Bandwidth Usage Reduction Vs. Accuracy (\%)}
\label{table_bw_opt}
\vskip -0.1in
\begin{tabular}{r|rr|rr}
	\hline
	{Method} & \multicolumn{2}{|c|}{{Original RGB}} & \multicolumn{2}{c}{{Our Compression}} \\
	\hline
	\rottt{Image Size} & \rottt{Accuracy} & \rottt{{BW Reduction~}} & \rottt{Accuracy} & \rottt{BW Reduction~} \\
	\hline
	$64{\times}64$ & $93.0\%$ & $30{\times}$ & $95.3\%$ & $90{\times}$ \\
	$64{\times}128$ & $95.7\%$ & $15{\times}$ & $96.7\%$ & $45{\times}$ \\
	$128{\times}128$ & $95.7\%$ & $7.5{\times}$ & $97\%$ & $22.5{\times}$ \\
	$256{\times}256$ & $97.2\%$ & $2.5{\times}$ & $97.2\%$ & $5.5{\times}$ \\
  $350{\times}350$ & $97.2\%$ & $1.0{\times}$ & $97.2\%$ & $3.0{\times}$ \\
	\hline 
\end{tabular}
\vskip -1cm
\end{table}

\section{Examples of Misrecognition}
\begin{figure}[h!]
    \subfigure[$53 \rightarrow 5.3$]{\includegraphics[width=0.45\columnwidth]{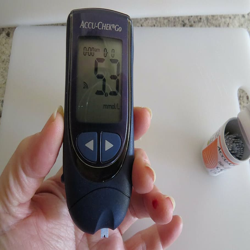}}
    \,
    \subfigure[$61 \rightarrow 67$]{\includegraphics[width=0.45\columnwidth]{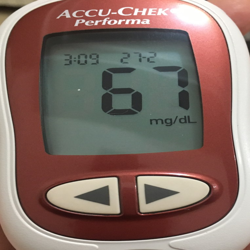}}
    \newline
    \subfigure[$5.81 \rightarrow 5.8$]{\includegraphics[width=0.45\columnwidth]{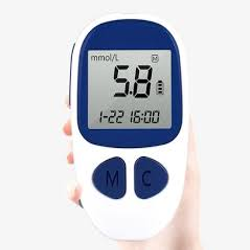}\label{mis_58}}
    \,
    \subfigure[$99 \rightarrow 399$]{\includegraphics[width=0.45\columnwidth]{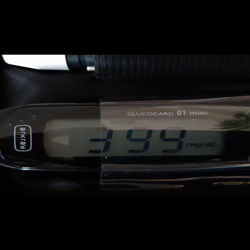}}
    \newline
    \subfigure[$94 \rightarrow 9.4$]{\includegraphics[width=0.45\columnwidth]{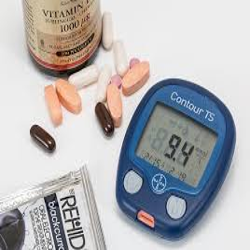}}
    \,
    \subfigure[$1067 \rightarrow 106$]{\includegraphics[width=0.45\columnwidth]{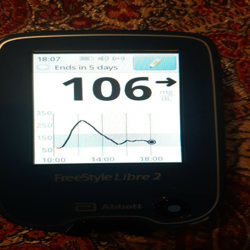}\label{mis_106}}
    \vskip -0.11in
    \caption{Examples of mispredicted images. (prediction $\rightarrow$ ground-truth)}
    \label{fig:mispredictions}
\end{figure}

We provide samples that our algorithm did not recognize them correctly. As shown in Fig\,\ref{fig:mispredictions}, in some situations, additional elements on the screen mislead our algorithm. For example, in Fig.\,\ref{mis_106} and Fig.\,\ref{mis_58} an arrow and battery information are wrongly detected as `7' and `1', respectively. This can be corrected by adding similar samples to the training set in the next iterations.

\end{document}